\documentclass[aps, prd,  showkeys, showpacs]{revtex4}

\usepackage[small]{caption}
\usepackage{amsmath,amsfonts,amscd,amssymb,epsf, epsfig}\usepackage{graphicx}

\newcommand{\Z}{{\mathbb{Z}}}

\newcommand{\R}{\mathbb{R}}

\newcommand{\pa}{\partial}

\begin{document}

\title{Finite temperature Casimir effect for massive scalar field in   spacetime with extra dimensions}

\author{L.P. Teo}\email{lpteo@mmu.edu.my}\affiliation{Faculty of Information
Technology, Multimedia University, Jalan Multimedia, Cyberjaya,
63100, Selangor Darul Ehsan, Malaysia.}

\keywords{Casimir effect, finite temperature, extra dimensions, massive scalar field, asymptotic behavior.}

\pacs{03.70.+k, 11.10.Kk    }

\begin{abstract}

We compute the finite temperature Casimir energy for massive scalar field with general curvature coupling subject to Dirichlet or Neumann boundary conditions on the walls of a closed cylinder with arbitrary cross section, located in a background spacetime of the form $M^{d_1+1}\times \mathcal{N}^n$, where $M^{d_1+1}$ is the $(d_1+1)$-dimensional Minkowski spacetime and $\mathcal{N}^n$ is an $n$-dimensional internal manifold. The Casimir energy is regularized using the criteria that it should vanish in the infinite mass limit. The Casimir force acting on a piston moving freely inside the closed cylinder is derived and it is shown that it is independent of the regularization procedure.  By letting one of the chambers of the cylinder divided by the piston to be infinitely long, we obtain the Casimir force acting on two parallel plates embedded in the cylinder. It is shown that if both the plates assume Dirichlet or Neumann boundary conditions, the strength of the Casimir force is reduced by the increase in mass. Under certain conditions, the passage from massless to massive will change the nature of the force from long range to short range. Other properties of the Casimir force such as its sign, its behavior at low and high temperature, and its behavior at small and large plate separations, are found to be similar to the massless case. Explicit exact formulas and asymptotic behaviors of the Casimir force at different limits are derived. The Casimir force when one plate assumes Dirichlet boundary condition and one plate assumes Neumann boundary condition is also derived and shown to be repulsive.

\end{abstract}
\maketitle

\section{ Introduction}

In the endeavor to solve some fundamental problems in physics, such as the unification of fundamental forces and the dark energy and cosmological constant problem, it has been proposed that we should consider higher dimensional spacetime. For example, string theory \cite{1} predicts that we live in spacetime of ten or eleven dimensions, where the extra six or seven space dimensions are curled up to a tiny invisible compact manifold. Therefore, there is a strong motivation to  study physics in spacetimes with extra dimensions.   Since Casimir effect is a fundamental quantum effect,   the influence of extra dimensions on Casimir effect becomes an important issue. The Casimir effect in spacetime with extra dimensions was considered in the works \cite{36, 37, 38, 3_17_1, 58, 59, 60, 46, 39, 56, 57, 77, 3_16_1, 2_10_1, 2_25_1, 3_17_2}. The role of Casimir effect in stabilizing extra dimensions was discussed in  \cite{48, 23, 52, 55, 41, 42, 43}. The possible role of Casimir effect as cosmological constant responsible for the observed dark energy was considered in \cite{28, 32, 66, 31, 39, 51, 3_17_3}. In \cite{38, 3_17_1}, the Casimir effect for massless scalar field in Kaluza-Klein spacetime of the form $M^{3+1}\times \mathcal{N}^n$, where $M^{3+1}$ is the $(3+1)$-dimensional Minkowski spacetime and $\mathcal{N}^n$ is an $n$-dimensional internal  manifold, was considered in the piston setting. In \cite{3_16_1, 2_10_1}, we extended the  results of \cite{38, 3_17_1} and considered the finite temperature correction to the Casimir effect. In the work \cite{2_25_1}, the Casimir effect due to massive scalar field with general curvature coupling constant subject to Robin boundary conditions in spacetime of the form $M^{d_1+1}\times \mathcal{N}^n$, where $M^{d_1+1}$ is the $(d_1+1)$-dimensional Minskowski spacetime, was considered.   In this paper, we generalize the work of \cite{2_25_1} by taking into account the temperature correction. However, we restrict ourselves to only consider either Dirichlet or Neumann boundary conditions, or  combinations of these two conditions. We study in detail both the effect of the mass and the temperature corrections to the Casimir effect.

We begin by computing the Casimir energy using exponential cut-off method when the massive scalar field is confined in a cylinder of arbitrary cross section in the background spacetime. Since a field should not have quantum fluctuations in the infinite mass limit \cite{2_19_1}, we impose the condition that the Casimir energy should vanish when the mass approaches infinity. We show that this condition allows us to regularize the Casimir energy and explicit formula of the regularized Casimir energy is given. We then consider the Casimir force in the piston setting \cite{2_16_1} -- a piston dividing a closed cylinder into two chambers --  which becomes fashionable nowadays. It is observed that the regularization procedure actually does not affect the Casimir force that acts on the piston. More precisely, the Casimir force acting on the piston is the same whether we compute it using the cut-off dependent Casimir energy before regularization or the regularized Casimir energy. The piston approach can be considered as the correct regularization procedure for computing Casimir force between two parallel plates \cite{38}. From the results for piston, we deduce the formula for the Casimir force acting on a pair of parallel plates embedded orthogonally in an infinitely long cylinder \cite{97}. By taking the limit where the cross section of the cylinder is infinitely large, we obtain the finite temperature Casimir force density acting on a pair of infinite parallel plates in a $(d_1+1)$-dimensional macroscopic Minkowski spacetime, with the presence of an $n$-dimensional internal manifold. An advantage of our approach is that we obtain a formula for the Casimir force acting on a pair of parallel plates embedded in an infinitely long cylinder as a series over elementary functions, which enables us to derive some properties of the Casimir force easily. It is shown that the Casimir force is attractive if both the plates assume Dirichlet boundary conditions or both plates assume Neumann boundary conditions. Moreover,   the magnitude of the Casimir force is always a decreasing function of the plate separation $a$ and the  mass $m$.  The latter supports the assumption that quantum fluctuations vanish in the infinite mass limit. In the case where the surrounding cylinder assumes Neumann boundary conditions, we show that taking the massless limit will change the nature of the Casimir force from short range to long range.
For the influence of the internal manifold, it is shown that the Casimir force is enhanced in the presence of extra dimensions. A stronger result shows that the Casimir force is an increasing function of the size of the internal manifold. By passing to the limit of infinite parallel plates, all these properties are preserved, although some of them is not obvious from the formulas for the case of  infinite parallel plates.

Besides the   properties of the Casimir force, we derive explicit formulas for the asymptotic behaviors of the Casimir force in different limits, such as low and high temperature, small plate separation, small mass and large cross section. It is shown that when the plate separation is small, the Casimir force is dominated by terms that are independent of mass. This shows that the effect of mass is less significant if the plate separation is small. On the other hand, the leading order term of the Casimir force is linear in temperature in the high temperature regime. In the case that the size of the internal manifold is comparable to the plate separations, the behavior of the Casimir force is quite complicated and it depends strongly on the geometry of the internal manifold.

Although we assume   that the macroscopic spacetime is Minkowskian, the results of this paper can be easily generalized to the case where the macroscopic spacetime is also curved. By setting the size of the internal manifold to be zero, one can obtain the corresponding results for spacetime without extra dimensions.

  Throughout this paper, we use the units where $\hbar=c=k_B=1$ except for the figures.

\section{Basic formalism}
We consider a background $(d+1)$-dimensional spacetime of the form $M^{d_1+1}\times \mathcal{N}^n$, where $M^{d_1+1}$ is the $(d_1+1)$-dimensional Minkowski spacetime,   $\mathcal{N}^n$ is an $n$-dimensional internal space, which is assumed to be a compact connected manifold without boundary and $d=d_1+n$.  Let the spacetime metric be given by
\begin{equation*}\begin{split}
ds^2 =& g_{\mu\nu} dx^{\mu}dx^{\nu} =\eta_{\alpha\beta}dx^{\alpha}dx^{\beta}-G_{ab}dy^a dy^b,\\
\mu,\nu=&0,1,\ldots,d;\;\alpha,\beta=0,1,\ldots, d_1; \;a,b=1,\ldots, n,\end{split}
\end{equation*}    where  $\eta_{\alpha\beta}=\text{diag}\,(1, -1,\ldots, -1)$,  $y^{a} = x^{d_1+a}$ for $a=1,\ldots,n $ and $G_{ab} dy^ady^b$ is a Riemannian metric on $\mathcal{N}^n$. In this article, we are interested in studying the finite temperature Casimir effect for a massive scalar field $\varphi(x)$ satisfying the equation of motion
\begin{equation}\label{eq2_10_1}
\left(\frac{1}{\sqrt{|g|}}\pa_{\mu}\sqrt{|g|}g^{\mu\nu}\pa_{\nu} +m^2 +\xi \mathfrak{R}\right)\varphi(x)=0,
\end{equation}where $\mathfrak{R}$ is the scalar curvature of the background spacetime and $\xi$ is a coupling constant --- $\xi=0$ corresponds to minimal coupling and $\xi = (d-1)/4d$ corresponds to conformal coupling. We assume that the field $\varphi(x)$ is confined in a cylinder of the form $\text{cyl}=[0, L]\times \Omega\times \mathcal{N}^n$,  and  satisfies Dirichlet or Neumann boundary conditions on the boundary of the cylinder. Here $\Omega$ is a simply connected domain in $\R^{d_1-1}$.

If the field $\varphi(x)$ satisfies Dirichlet boundary conditions, the eigenfunctions satisfying \eqref{eq2_10_1} are given by
\begin{equation}\label{eq2_10_2}
\varphi_{k,j,l}(x) = e^{-i\omega_{k,j,l} t} \sin \frac{\pi k x^1}{L}\phi_{  D; j}(x^2, \ldots, x^{d_1}) \Phi_{l}(y),
\end{equation}where $k, j\in \mathbb{N}$, $l\in \tilde{\mathbb{N}}=\mathbb{N}\cup\{0\}$. The function $\phi_{D; j}(x^2, \ldots, x^{d_1})$ is an eigenfunction of the Laplace operator  with Dirichlet boundary conditions on $\Omega$, i.e.,
\begin{equation*}\begin{split}
-\sum_{i=2}^{d_1}\frac{\pa^2}{\pa (x^i)^2} \phi_{D;j}(x^2, \ldots, x^{d_1})= &\omega_{\Omega, D; j}^2 \phi_{D;j}(x^2, \ldots, x^{d_1}),\\\;\;\left.\phi_{D;j}\right|_{\pa\Omega}=&0.\end{split}
\end{equation*} The function $\Phi_l(y)$ is an eigenfunction of the   operator $-\Delta_G-\xi \mathfrak{R}_G=-\sqrt{G}^{-1} \pa_{a} \sqrt{G} G^{ab} \pa_b - \xi \mathfrak{R}_G $, ($\mathfrak{R}_G=-\mathfrak{R}$ is the scalar curvature of the metric $G_{ab}dy^ady^b$) on $\mathcal{N}^n$ with eigenvalue $\omega_{\mathcal{N}; l}^2$, i.e.,
\begin{equation*}
-\left(\Delta_G+\xi \mathfrak{R}_G\right)\Phi_l(y) = \omega_{\mathcal{N}; l}^2 \Phi_l(y).
\end{equation*}The eigenfrequency $\omega_{k,j,l}$ is  given by
\begin{equation*}
\omega_{k,j,l}=\sqrt{\left(\frac{\pi k}{L}\right)^2+\omega_{\Omega, D;j}^2+\omega_{\mathcal{N}; l}^2+m^2}.
\end{equation*}  Since $\Omega$ is simply connected, the eigenvalues $\omega_{\Omega, D;j}^2$ are all nonzero.  For Neumann boundary conditions, the eigenfunctions are
\begin{equation}\label{eq2_10_3}
\varphi_{k,j,l}(x) = e^{-i\omega_{k,j,l} t} \cos \frac{\pi k x^1}{L}\phi_{  N; j}(x^2, \ldots, x^{d_1}) \Phi_{l}(y),
\end{equation}where   $k,j,l\in \tilde{\mathbb{N}}=\mathbb{N}\cup\{0\}$. The function $\phi_{N; j}(x^2, \ldots, x^{d_1})$ is an eigenfunction of the Laplace operator  with Neumann boundary conditions on $\Omega$, i.e.,
\begin{equation*}\begin{split}
-\sum_{i=2}^{d_1}\frac{\pa^2}{\pa (x^i)^2} \phi_{N;j}(x^2, \ldots, x^{d_1})= &\omega_{\Omega, N; j}^2 \phi_{N;j}(x^2, \ldots, x^{d_1}),\\\;\;\left.\frac{\pa \phi_{N;j}}{\pa \mathbf{n}}\right|_{\pa\Omega}=&0,\end{split}
\end{equation*}where $\mathbf{n}$ is a unit vector perpendicular to $\pa\Omega$.    By convention, $\phi_{N;0}(x^2, \ldots, x^{d_1})$ is the constant function with zero eigenvalue. For the eigenvalues $\omega_{\mathcal{N};l}^2$, we assume that $\omega_{\mathcal{N};l}^2\geq 0$ and there are exactly $\kappa$ of them that are equal to zero.

The boundary conditions considered above are homogeneous  boundary conditions. In the following, we are going to consider the Casimir force for parallel plates and also Casimir force in the piston setup. In these scenarios, we will also consider   mixed boundary conditions. We consider two cases. For the first case, we assume that  the field $\varphi(x)$ satisfies Dirichlet boundary conditions on the wall $[0, L]\times \pa\Omega\times \mathcal{N}^n$ and the wall $x^1=0$, and Neumann boundary condition on the wall $x^1=L$.  The  eigenfunctions are obtained from \eqref{eq2_10_2} by replacing the function $\sin \frac{\pi kx}{L}$ by $\sin \frac{\pi \left(k-\frac{1}{2}\right)x}{L}$. For the second type, we interchange the roles of   Dirichlet and Neumann on the boundaries. In this case, the eigenfunctions are obtained from \eqref{eq2_10_3} by replacing $\cos\frac{\pi k x}{L}$ by $\cos\frac{\pi \left(k+\frac{1}{2}\right)x}{L}$.

Before ending this section, we define the variables $R$ and $r$ by $R= \text{Vol}(\Omega)^{1/(d_1-1)}$ and $r= \text{Vol}(\mathcal{N})^{1/n}$ which have the dimension of length to measure the size of the domain $\Omega$ and the internal manifold $\mathcal{N}^n$.  Throughout this article, we assume that the size $r$ of the internal manifold is smaller than any measurable length in the $(d_1+1)$-dimensional Minkowski spacetime. The re-scaled variables \begin{equation*}\begin{split}\omega_{\Omega, * ;j}'=&R\omega_{\Omega,*;j},\;\;\;*=D\;\text{or}\;N,\\
\omega_{\mathcal{N}; l}'=&r\omega_{\mathcal{N};l},\end{split}\end{equation*}are  dimensionless variables and are invariant under the re-scaling of the domain $\Omega$ and the manifold $\mathcal{N}^n$.

\section{The Casimir energy}

\subsection{Zero temperature Casimir energy}
At zero temperature, the Casimir energy is defined naively as the sum of zero point energies:
\begin{equation*}
E_{\text{Cas}}^{T=0} (L) = \frac{1}{2}\sum \omega_{k,j,l}.
\end{equation*}To take into account the homogeneous and mixed boundary conditions, we let
\begin{equation*}\begin{split}
\omega_{k,j, l}=&\omega_{k,j, l}(\alpha,*;m) = \sqrt{\left(\frac{\pi (k+\alpha)}{L}\right)^2+\omega_{\Omega, *, j}^2+\omega_{\mathcal{N};l}^2+m^2},  \hspace{1cm}  k\in \tilde{\mathbb{N}}, j\in J_*, l\in \tilde{\mathbb{N}},\end{split}
\end{equation*}where $\alpha=1$ for homogeneous Dirichlet boundary conditions, $\alpha=0$ for homogeneous Neumann boundary conditions and $\alpha=1/2$ for mixed boundary conditions; $*=D$ (resp. $*=N$) when $\varphi(x)$ assume Dirichlet (resp. Neumann) boundary conditions on the component   $[0,L]\times\pa\Omega\times\mathcal{N}^n$ of the boundary of the cylinder; $J_D=\mathbb{N}$ and $  J_N=\tilde{\mathbb{N}}$. Using exponential cut-off regularization, we define the cut-off dependent energy as
\begin{equation*}
E_{\text{Cas}}^{T=0} (L;\lambda) = \frac{1}{2}\sum_{k=0}^{\infty}\sum_{j\in J_*}\sum_{l=0}^{\infty} \omega_{k,j,l}e^{-\lambda\omega_{k,j,l}}.
\end{equation*}Using the same method as the massless case \cite{2_10_1}, we find that up to the term $\lambda^0$, the small-$\lambda$ expansion of $E_{\text{Cas}}^{T=0} (L;\lambda)$ is given by
\begin{equation}\label{eq2_10_11}\begin{split}
&E_{\text{Cas}}^{T=0} (L;\lambda) =\sum_{i=0}^{d-1} \frac{\Gamma(d+1-i)}{\Gamma\left(\frac{d-i}{2}\right)}{c_{\text{cyl},\alpha,*; i}(m)}{\lambda^{i-d-1}} -\frac{\psi(1)-\log \lambda}{2\sqrt{\pi}}c_{\text{cyl},\alpha,*; d+1}(m)+\frac{1}{2}\text{FP}_{s=-\frac{1}{2}}\zeta_{\text{cyl}, \alpha,*}(s; m), \end{split}
\end{equation}where $c_{\text{cyl}, \alpha, *; i}(m)$ are the heat kernel coefficients defined so that
\begin{equation}\label{eq2_11_6}
\begin{split}&\sum_{k=0}^{\infty} \sum_{j\in J_*}\sum_{l=0}^{\infty} e^{ -t\omega_{k,j,l}^2}
=\sum_{i=0}^{M-1} c_{\text{cyl}, \alpha, *; i}(m) t^{\frac{i-d}{2}}+O\left(t^{\frac{M-d}{2}}\right)\hspace{0.5cm}\text{as}\;\;t\rightarrow 0^+,\end{split}
\end{equation}$\zeta_{\text{cyl}, \alpha,*}(s; m)$ is the zeta function
\begin{equation*}
\zeta_{\text{cyl}, \alpha,*}(s; m)=\sum_{k=0}^{\infty} \sum_{j\in J_*}\sum_{l=0}^{\infty}\omega_{k,j,l}^{-2s},
\end{equation*}and the finite part of a meromorphic function $f(z)$ with at most simple pole at a point $z=z_0$ is defined by
\begin{equation*}
\text{FP}_{z=z_0} f(z) = \lim_{z\rightarrow z_0}\left( f(z)-\frac{\text{Res}_{z=z_0}f(z)}{z-z_0}\right).
\end{equation*}As $\lambda\rightarrow 0^+$, we see from \eqref{eq2_10_11} that the divergent part of the Casimir energy contains divergence of order $\log\lambda$ and  $\lambda^{-i}$, $i=2, \ldots, d+1$, with coefficients depending on the coefficients $c_{\text{cyl}, \alpha, *; i}(m)$, which can be expressed in terms of $L$, $m$ and the geometric invariants of the manifolds $\Omega$ and $\mathcal{N}^n$. In particular, the leading divergence
\begin{equation*}
\frac{\Gamma(d+1)}{\Gamma\left(\frac{d}{2}\right)} c_{\text{cyl}, \alpha, *; 0}(m)\lambda^{-d-1}=\frac{\Gamma(d+1)}{\Gamma\left(\frac{d}{2}\right)}\frac{L\text{Vol}(\Omega)\text{Vol}(\mathcal{N}^n)}{2^d\pi^{\frac{d}{2}}}\lambda^{-d-1}
\end{equation*}  is called the bulk divergence and is usually subtracted away in the definition of Casimir energy. The other divergences are called hypersurface divergences and regularization is required to remove these divergences. A conventional method, known as zeta regularization \cite{2_11_1}, set all the hypersurface divergences to zero and define the regularized zero temperature Casimir energy to be
\begin{equation}\label{eq2_11_1}\begin{split}
E_{\text{Cas}}^{\text{reg}_{\zeta},T=0} (L) =&\frac{1}{2}\text{FP}_{s=-\frac{1}{2}}\zeta_{\text{cyl}, \alpha,*}(s; m)+\frac{1}{2}\log\mu^2\text{Res}_{s=-\frac{1}{2}}\zeta_{\text{cyl}, \alpha,*}(s; m)\\=&\frac{1}{2}\text{FP}_{s=-\frac{1}{2}}\zeta_{\text{cyl}, \alpha,*}(s; m)-\frac{\log\mu}{2\sqrt{\pi}}c_{\text{cyl},\alpha,*; d+1}(m),
\end{split}\end{equation}where $\mu$ is a normalization constant with dimension length$^{-1}$. This is tantamount to subtracting   the divergent terms
\begin{equation}\label{eq2_13_1}\begin{split}
&\sum_{i=0}^{d-1} \frac{\Gamma(d+1-i)}{\Gamma\left(\frac{d-i}{2}\right)}{c_{\text{cyl},\alpha,*; i}(m)}{\lambda^{i-d-1}}-\frac{\psi(1)-\log (\lambda\mu)}{2\sqrt{\pi}}c_{\text{cyl},\alpha,*; d+1}(m)
\end{split}\end{equation}from the cut-off dependent Casimir energy \eqref{eq2_10_11} and set $\lambda=0$.
However, for massive scalar fields, this definition is not sufficient. Since a field should not have quantum fluctuations in the limit of infinite mass \cite{2_19_1},    it  is natural to require   the regularized Casimir energy $E_{\text{Cas}}^{\text{reg}}$  to vanish when the mass $m$ approaches infinity, i.e.,
\begin{equation*}
E_{\text{Cas}}^{\text{reg}} \xrightarrow{m\rightarrow \infty} 0.
\end{equation*} To satisfy this condition, one need to extract the leading behavior of the Casimir energy in \eqref{eq2_10_11} or \eqref{eq2_11_1} when the mass is large and subtract away those terms that give nontrivial limits when the mass tends to infinity. Using the fact that
\begin{equation*}\begin{split}
&\sum_{k=0}^{\infty} \sum_{j\in J_*}\sum_{l=0}^{\infty} e^{ -t\omega_{k,j,l}^2}=\sum_{k=0}^{\infty} \sum_{j\in J_*}\sum_{l=0}^{\infty}e^{-tm^2}\exp\left\{ -t\left(\left[\frac{\pi(k+\alpha)}{L}\right]^2+\omega_{\Omega, *,j}^2+\omega_{\mathcal{N}, l}^2\right)\right\},
\end{split}\end{equation*} it is easy to show that the heat kernel coefficients $c_{\text{cyl},\alpha,*; i}(m)$, $i=0,1,2,\ldots$ at any mass can be expressed as a polynomial in $m^2$ with coefficients in terms of the heat kernel coefficients  $c_{\text{cyl},\alpha,*; d+1}(0)$ when $m=0$. More precisely,
\begin{equation*}
c_{\text{cyl},\alpha,*; i}(m)=\sum_{j=0}^{\left[\frac{i}{2}\right]}\frac{(-1)^j}{j!} c_{\text{cyl},\alpha,*; i-2j}(0)m^{2j}.
\end{equation*}Therefore, the divergent part of the Casimir energy \eqref{eq2_13_1}   can be expressed as a polynomial in $m^2$. As a result, it has to be subtracted away in the regularization procedure and therefore do not contribute to the regularized Casimir energy. For the regular part of the Casimir energy given by \eqref{eq2_11_1}, the second term that proportional to $c_{\text{cyl},\alpha,*; d+1}(m)$ would not contribute to the regularized Casimir energy for the same reason above.
For the term $(1/2)\text{FP}_{s=-\frac{1}{2}}\zeta_{\text{cyl}, \alpha,*}(s; m)$, we write the zeta function $\zeta_{\text{cyl}, \alpha,*}(s; m)$ as the Mellin transform of a heat kernel
\begin{equation*}
\zeta_{\text{cyl}, \alpha,*}(s; m)=\frac{1}{\Gamma(s)}\int_0^{\infty} t^{s-1} \sum_{k=0}^{\infty} \sum_{j\in J_*}\sum_{l=0}^{\infty} e^{ -t\omega_{k,j,l}^2}dt.
\end{equation*}Define
\begin{equation*}\begin{split}
\tilde{K}(t) = \sum_{k=0}^{\infty} \sum_{j\in J_*}\sum_{l=0}^{\infty} \exp\left\{ -t\left(\left[\frac{\pi(k+\alpha)}{L}\right]^2+\omega_{\Omega, *,j}^2+\omega_{\mathcal{N}, l}^2\right)\right\} -\sum_{i=0}^{d+1} c_{\text{cyl}, \alpha, *; i}(0) t^{\frac{i-d}{2}}.
\end{split}\end{equation*}Then
\begin{equation*}
\begin{split}
\tilde{K}(t)=&O\left(t\right) \;\;\text{as}\;\; t\rightarrow 0^+,\hspace{1cm}
\tilde{K}(t)=O\left(t^{\frac{1}{2}}\right) \;\;\text{as}\;\; t\rightarrow \infty,
\end{split}
\end{equation*}
and we find that
\begin{equation*}\begin{split}
\zeta_{\text{cyl}, \alpha,*}(s; m)=&\frac{1}{\Gamma(s)}\int_0^{\infty} t^{s-1} e^{-tm^2}\left(\sum_{i=0}^{d+1} c_{\text{cyl}, \alpha, *; i}(0) t^{\frac{i-d}{2}}+\tilde{K}(t)\right)dt\\
=& \sum_{i=0}^{d+1} \frac{\Gamma\left(s+\frac{i-d}{2}\right)}{\Gamma(s)}c_{\text{cyl}, \alpha, *; i}(0) m^{-2s-i+d} +\frac{1}{\Gamma(s)}\int_0^{\infty} t^{s-1} e^{-tm^2}\tilde{K}(t)dt
\end{split}\end{equation*}gives an analytic continuation of $\zeta_{\text{cyl}, \alpha,*}(s; m)$ into the domain $\text{Re}\;s>-1$. Putting $s=-1/2$, we find that
\begin{equation}\label{eq2_13_2}\begin{split}
\frac{1}{2}\text{FP}_{s=-\frac{1}{2}}\zeta_{\text{cyl}, \alpha,*}(s; m)=&\frac{\psi\left(-\frac{1}{2}\right)+\log m^2}{4\sqrt{\pi}}c_{\text{cyl},\alpha,*;d+1}(m)
+\mathfrak{D}(m)-\frac{1}{4\sqrt{\pi}}\int_0^{\infty} t^{-\frac{3}{2}}
e^{-tm^2}\tilde{K}(t)dt,\end{split}\end{equation}where \begin{equation*}
\mathfrak{D}(m) =
-\frac{1}{4\sqrt{\pi}}\sum_{i=0}^{d+1}c_{\text{cyl},\alpha,*;d+1-i}(0)m^i\text{FP}_{s=-\frac{i}{2}}\Gamma(s).
\end{equation*} Therefore, as $m\rightarrow \infty$, \begin{equation*}\begin{split}
\frac{1}{2}&\text{FP}_{s=-\frac{1}{2}}\zeta_{\text{cyl}, \alpha,*}(s; m)=\frac{\psi\left(-\frac{1}{2}\right)+\log m^2}{4\sqrt{\pi}}c_{\text{cyl},\alpha,*;d+1}(m)
+\mathfrak{D}(m) +o(1).\end{split}\end{equation*}This implies that to obtain the regularized Casimir energy that vanishes when $m\rightarrow \infty$, we should subtract away the term
\begin{equation}\label{eq2_16_5}
\begin{split}
&\sum_{i=0}^{d-1} \frac{\Gamma(d+1-i)}{\Gamma\left(\frac{d-i}{2}\right)}{c_{\text{cyl},\alpha,*; i}(m)}{\lambda^{i-d-1}}+\frac{\psi\left(-\frac{1}{2}\right)-2\psi(1)+\log [\lambda m]^2}{4\sqrt{\pi}}c_{\text{cyl},\alpha,*; d+1}(m)+\mathfrak{D}(m)
\end{split}
\end{equation}from the cut-off dependent Casimir energy \eqref{eq2_10_11} or subtract away the term
\begin{equation*}\begin{split}
&\frac{\psi\left(-\frac{1}{2}\right)+\log \left(m/\mu\right)^2}{4\sqrt{\pi}}c_{\text{cyl},\alpha,*;d+1}(m)
+\mathfrak{D}(m)
\end{split}\end{equation*}from the zeta regularized Casimir energy  \eqref{eq2_11_1}. From \eqref{eq2_13_2}, we then find that the regularized Casimir energy is given by
\begin{equation}\label{eq3_19_1}\begin{split}
E_{\text{Cas}}^{\text{reg} ,T=0} (L)=&\frac{1}{2}\text{FP}_{s=-\frac{1}{2}}\zeta_{\text{cyl}, \alpha,*}(s; m)-\frac{\psi\left(-\frac{1}{2}\right)+\log m^2}{4\sqrt{\pi}} c_{\text{cyl},\alpha,*;d+1}(m)
-\mathfrak{D}(m)=-\frac{1}{4\sqrt{\pi}}\int_0^{\infty} t^{-\frac{3}{2}} e^{-tm^2}\tilde{K}(t)dt.
\end{split}\end{equation}To study the behavior of the regularized Casimir energy with respect to the variation of the plate separation $L$, we use the fact that for  $\alpha=0,1/2, 1$,
\begin{equation}\label{eq2_10_5}\begin{split}
&\sum_{k=0}^{\infty} e^{-t \frac{\pi^2\left(k+\alpha\right)^2}{L^2}}=\frac{1}{2}\sum_{k=-\infty}^{\infty} e^{-t \frac{\pi^2\left(k+\alpha\right)^2}{L^2}}+\frac{1}{2}-\alpha
=\frac{L}{2\sqrt{\pi}}\sum_{k=-\infty}^{\infty} t^{-\frac{1}{2}}e^{2\pi i k \alpha} e^{-\frac{k^2 L^2}{t}} +\frac{1}{2}-\alpha. \end{split}
\end{equation}This implies that
\begin{equation}\label{eq2_13_4}\begin{split}
\zeta_{\text{cyl}, \alpha,*}(s; m)=&\frac{1}{\Gamma(s)}\int_0^{\infty} t^{s-1}  \sum_{j\in J_*}\sum_{l=0}^{\infty}e^{-t(\omega_{\Omega, *,j}^2+\omega_{\mathcal{N};l}^2+m^2)}   \left\{\frac{L}{2\sqrt{\pi}}\sum_{k=-\infty}^{\infty} t^{-\frac{1}{2}}e^{2\pi i k \alpha} e^{-\frac{k^2 L^2}{t}} +\frac{1}{2}-\alpha\right\}dt\\
=&\left(\frac{1}{2}-\alpha\right)\zeta_{\Omega\times\mathcal{N}, *}(s; m) +\frac{L}{2\sqrt{\pi}}\frac{\Gamma\left(s-\frac{1}{2}\right)}{\Gamma(s)}\zeta_{\Omega\times\mathcal{N}, *}\left(s-\frac{1}{2}; m\right)\\
&+\frac{2L}{\sqrt{\pi}\Gamma(s)}\sum_{k=1}^{\infty}\sum_{j\in J_*}\sum_{l=0}^{\infty} e^{2\pi i k\alpha} \left(\frac{kL}{\sqrt{\omega_{\Omega, *,j}^2+\omega_{\mathcal{N};l}^2+m^2}}\right)^{s-\frac{1}{2}}  K_{s-\frac{1}{2}}\left( 2kL\sqrt{\omega_{\Omega, *,j}^2+\omega_{\mathcal{N};l}^2+m^2} \right),\end{split}
\end{equation}where $K_{\nu}(z)$ is the modified Bessel function of second kind, and
\begin{equation*}
\zeta_{\Omega\times\mathcal{N}, *}(s; m) =\sum_{j\in J_*}\sum_{l=0}^{\infty} \left(\omega_{\Omega, *,j}^2+\omega_{\mathcal{N};l}^2+m^2\right)^{-s}.
\end{equation*}As a result, we have
\begin{equation}\label{eq2_13_3}
\begin{split}
&E_{\text{Cas}}^{\text{reg} ,T=0} (L)=\frac{1-2\alpha}{4} \text{FP}_{s=-\frac{1}{2}}\zeta_{\Omega\times\mathcal{N}, *}(s; m)
+\frac{L}{8\pi}\Biggl\{ \psi\left(-\frac{1}{2}\right) \text{Res}_{s=-1}\left(\Gamma(s)\zeta_{\Omega\times\mathcal{N}, *}(s)\right) - \text{FP}_{s=-1}\left(\Gamma(s)\zeta_{\Omega\times\mathcal{N}, *}(s)\right)\Biggr\}\\&-\frac{\psi\left(-\frac{1}{2}\right)+\log m^2}{4\sqrt{\pi}}   c_{\text{cyl},\alpha,*;d+1}(m)
-\mathfrak{D}(m)-\frac{1}{2\pi}\sum_{k=1}^{\infty}\sum_{j\in J_*}\sum_{l=0}^{\infty} e^{2\pi i k\alpha} \frac{\sqrt{\omega_{\Omega, *,j}^2+\omega_{\mathcal{N};l}^2+m^2} }{k}K_1\left(2kL \sqrt{\omega_{\Omega, *,j}^2+\omega_{\mathcal{N};l}^2+m^2} \right).\end{split}
\end{equation}Using \eqref{eq2_10_5}, we also have
\begin{equation*}
\begin{split}\sum_{k=0}^{\infty} \sum_{j\in J_*}\sum_{l=0}^{\infty} e^{ -t\omega_{k,j,l}^2}=&\left(\frac{L}{2\sqrt{\pi}}t^{-\frac{1}{2}}+\frac{1}{2}-\alpha+\text{e.d.} \right)
\sum_{j\in J_*}\sum_{l=0}^{\infty} e^{-t\left(\omega_{\Omega, *,j}^2+\omega_{\mathcal{N}, l}^2+m^2\right)}
\\=&\left(\frac{L}{2\sqrt{\pi}}t^{-\frac{1}{2}}+\frac{1}{2}-\alpha+\text{e.d.}\right) \left(\sum_{i=0}^{M-1} c_{\Omega\times\mathcal{N}, *;i}(m)t^{\frac{i-d+1}{2}}+O\left( t^{\frac{M-d+1}{2}}\right)\right)\;\;\text{as}\;\;t\rightarrow 0^+, \end{split}
\end{equation*}where e.d. is the exponentially decay  terms, and $c_{\Omega\times\mathcal{N}, *;i}(m)$ are heat kernel coefficients for an elliptic operator on $\Omega\times \mathcal{N}$. This  implies that
\begin{equation}\label{eq2_16_9}
c_{\text{cyl}, \alpha,*; i}(m)=\frac{L}{2\sqrt{\pi}} c_{\Omega\times\mathcal{N}, *;i}(m)+\left(\frac{1}{2}-\alpha\right)c_{\Omega\times\mathcal{N}, *;i-1}(m).
\end{equation}Therefore, we read from \eqref{eq2_13_3} that the regularized Casimir energy can be written as a linear polynomial in $L$, plus a term that decays exponentially as $L$ approaches infinity, i.e.,
\begin{equation}\label{eq2_16_1}\begin{split}
&E_{\text{Cas}}^{\text{reg} ,T=0} (L)=\Sigma_0 +\Sigma_1 L -\frac{1}{2\pi}\sum_{k=1}^{\infty}\sum_{j\in J_*}\sum_{l=0}^{\infty} e^{2\pi i k\alpha} \frac{\sqrt{\omega_{\Omega, *,j}^2+\omega_{\mathcal{N};l}^2+m^2} }{k}K_1\left(2kL \sqrt{\omega_{\Omega, *,j}^2+\omega_{\mathcal{N};l}^2+m^2} \right).
\end{split}\end{equation}

\subsection{Finite temperature Casimir energy}
Now we take into account the temperature correction to the Casimir energy given by
\begin{equation}\label{eq2_16_3}
\begin{split}
T\sum \log\left(1-e^{-\omega_{k,j,l}/T}\right).
\end{split}
\end{equation}This summation is finite and no regularization is required. Therefore, the cut-off dependent finite temperature Casimir energy is defined as
\begin{equation*}
\begin{split}
E_{\text{Cas}}  (L;\lambda) = &\frac{1}{2}\sum_{k=0}^{\infty}\sum_{j\in J_*}\sum_{l=0}^{\infty} \omega_{k,j,l}e^{-\lambda\omega_{k,j,l}}+T\sum \log\left(1-e^{-\omega_{k,j,l}/T}\right).
\end{split}
\end{equation*}
It is well known that this  can be computed  in terms of the finite temperature zeta function given by
\begin{equation*}
\zeta_{\text{cyl}, \alpha, *; T}(s;m) = \sum_{k=0}^{\infty} \sum_{j\in J_*}\sum_{l=0}^{\infty} \sum_{p=-\infty}^{\infty}e^{ -t\left(\omega_{k,j,l}^2+(2\pi p T)^2\right)}.
\end{equation*}More precisely (see \cite{2_10_1}),
\begin{equation}\label{eq2_16_4}\begin{split}
E_{\text{Cas}}  (L;\lambda) = &\sum_{i=0}^{d-1} \frac{\Gamma(d+1-i)}{\Gamma\left(\frac{d-i}{2}\right)}{c_{\text{cyl},\alpha,*; i}(m)}{\lambda^{i-d-1}}+\frac{\log [\lambda\mu]-\psi(1)-\log2+1}{2\sqrt{\pi}}c_{\text{cyl},\alpha,*; d+1}(m)\\&-\frac{T}{2}\left(\zeta_{\text{cyl}, \alpha, *; T}'(0;m)+\log[\mu^2]\zeta_{\text{cyl}, \alpha, *; T}(0;m)\right).
\end{split}\end{equation}For the regularization, we observe that the temperature correction to the Casimir energy \eqref{eq2_16_3} vanishes when the mass $m$ approach infinity. Therefore, the large-$m$ leading behavior of the cut-off dependent finite temperature Casimir energy \eqref{eq2_16_4} is the same as the large-$m$ leading behavior of  the cut-off dependent zero temperature Casimir energy given by \eqref{eq2_16_5}. After subtracting away these large-$m$ non-vanishing terms, we find that the regularized finite temperature Casimir energy is given by
\begin{equation}\label{eq2_16_7}\begin{split}
E_{\text{Cas}}^{\text{reg}}  (L ) =&  -\frac{T}{2}\left(\zeta_{\text{cyl}, \alpha, *; T}'(0;m)+\log[\mu^2]\zeta_{\text{cyl}, \alpha, *; T}(0;m)\right) -\frac{\log \left[\frac{m}{\mu}\right]^2+\psi(1) }{4\sqrt{\pi}}c_{\text{cyl},\alpha,*; d+1}(m)-\mathfrak{D}(m).
\end{split}\end{equation}As a remark, in the zeta regularization scheme, the regularized finite temperature Casimir energy is defined as
\begin{equation}\label{eq2_16_8}\begin{split}
&E_{\text{Cas}}^{\text{reg}_{\zeta}}  (L ) = -\frac{T}{2}\left(\zeta_{\text{cyl}, \alpha, *; T}'(0;m)+\log[\mu^2]\zeta_{\text{cyl}, \alpha, *; T}(0;m)\right).
\end{split}\end{equation}This  expression does not go to zero as the mass $m$ approaches infinity. However, if we subtract away from \eqref{eq2_16_8} the large-$m$ leading terms that give nontrivial limits, we would   obtain \eqref{eq2_16_7}. We would like to remark that although the expression \eqref{eq2_16_7} for the regularized Casimir energy is derived for the case where the field is confined within a cylindrical cavity, this formula is in fact valid for an arbitrary closed cavity. Similarly, the formula for the regularized Casimir energy at zero temperature given by \eqref{eq3_19_1} is also valid for any closed cavity. We would also like to point out that the high temperature expansion of the zeta regularized finite temperature Casimir energy for massive scalar field in manifolds with boundaries have been obtained in \cite{1_17_1, 1_5_1, 1_17_2, 1_17_3, 79}.

 Using the same method as in the derivation of \eqref{eq2_13_4} (see \cite{2_10_1}), we find that
 \begin{equation*}\begin{split}
  \zeta_{\text{cyl},\alpha,*;T}(0;m)=&\frac{c_{\text{cyl},\alpha,*; d+1}(m)}{2\sqrt{\pi}T},
 \end{split}\end{equation*}\begin{equation*}\begin{split}
 \zeta_{\text{cyl},\alpha,*;T}'(0;m)=&\Lambda_0+\Lambda_1 L +  \sum_{k=1}^{\infty}\sum_{j\in J_*}\sum_{l=0}^{\infty} \sum_{p=-\infty}^{\infty}\frac{e^{2\pi i k\alpha}}{k}  \exp\left(-2kL\sqrt{\omega_{\Omega, *,j}^2+\omega_{\mathcal{N}, l}^2+m^2+(2\pi p T)^2}\right),
\end{split}
\end{equation*}where $\Lambda_0$ and $\Lambda_1$ are independent of $L$, and $\Lambda_1$ is given by
\begin{equation*}
\begin{split}
\Lambda_1=&-\frac{\psi(1)}{4\pi T}c_{\Omega\times \mathcal{N}, *; d+1}(m) +\frac{1}{4\pi T}\text{FP}_{s=-1}\left\{\Gamma(s)\zeta_{\Omega\times \mathcal{N}}(s; m)\right\}\\&+\frac{2}{\pi}\sum_{j\in J_*}\sum_{l=0}^{\infty}\sum_{p=1}^{\infty}
\frac{\sqrt{\omega_{\Omega,*; j}^2+\omega_{\mathcal{N};l}^2+m^2}}{p}    K_1\left(\frac{p\sqrt{\omega_{\Omega,*; j}^2+\omega_{\mathcal{N};l}^2+m^2}}{T}\right).
\end{split}
\end{equation*}Together with \eqref{eq2_16_9}, we find that the regularized Casimir energy \eqref{eq2_16_7} can be expressed as the sum of a linear polynomial in $L$ plus a term that decays exponentially as $L$ approaches infinity, i.e.,
\begin{equation}\label{eq2_16_10}\begin{split}
&E_{\text{Cas}}^{\text{reg}  } (L)=\tilde{\Sigma}_0 +\tilde{\Sigma}_1 L -\frac{T}{2}\sum_{k=1}^{\infty}\sum_{j\in J_*}\sum_{l=0}^{\infty} \sum_{p=-\infty}^{\infty}\frac{e^{2\pi i k\alpha}}{k}  \exp\left(-2kL\sqrt{\omega_{\Omega, *,j}^2+\omega_{\mathcal{N}, l}^2+m^2+(2\pi p T)^2}\right),
\end{split}\end{equation}where
\begin{equation}\label{eq2_18_1}
\begin{split}
\tilde{\Sigma}_1= &\frac{1}{8\pi}\sum_{i=0}^{d+1}\left[\text{FP}_{s=-\frac{i}{2}}\Gamma(s)\right] c_{\Omega\times\mathcal{N},*;d+1-i}(0)m^i -\frac{1}{8\pi}\left\{\text{FP}_{s=-1}\left\{\Gamma(s)\zeta_{\Omega\times \mathcal{N}}(s; m)\right\}+\log [m^2] c_{\Omega\times\mathcal{N},*;d+1}(m)\right\}\\
&-\frac{T}{\pi}\sum_{j\in J_*}\sum_{l=0}^{\infty}\sum_{p=1}^{\infty}
\frac{\sqrt{\omega_{\Omega,*; j}^2+\omega_{\mathcal{N};l}^2+m^2}}{p}   K_1\left(\frac{p\sqrt{\omega_{\Omega,*; j}^2+\omega_{\mathcal{N};l}^2+m^2}}{T}\right).
\end{split}
\end{equation}

Before ending this section, we would like to comment that it is easy to deduce from the results above that the Casimir energy for mixed boundary conditions ($\alpha=1/2)$)  are related to the Casimir energy for homogeneous boundary conditions ($\alpha=0$ or $1$) by
\begin{equation}\label{eq3_6_1}\begin{split}
E_{\text{Cas}}\left(L; \alpha=\frac{1}{2}, *= D\right) = & E_{\text{Cas}}\left(2L; \alpha=1, *= D\right)-E_{\text{Cas}}\left(L; \alpha=1, *= D\right),\\
E_{\text{Cas}}\left(L; \alpha=\frac{1}{2}, *= N\right) = & E_{\text{Cas}}\left(2L; \alpha=0, *= N\right)-E_{\text{Cas}}\left(L; \alpha=0, *= N\right).
\end{split}\end{equation}

\subsection{Special Case}\label{s3_3}
Here we consider the special case where the cross section of the cylinder $\Omega$ is a rectangular region $[0, L_2]\times \ldots\times [0, L_{d_1}]$ and the internal manifold $\mathcal{N}^n$ is an $n$-torus $T^n$ -- a product of $n$ circles with radius $r_1, \ldots, r_n$ respectively. Notice that the scalar curvature of the torus $T^n$ is zero. The spectrum $\omega_{\Omega, *; j}^2$ of the Laplace operator with Dirichlet or Neumann boundary conditions on $\Omega$ is given by
\begin{equation*}
\left(\frac{\pi j_2}{L_2}\right)^2 + \ldots +\left(\frac{\pi j_{d_1}}{L_{d_1}}\right)^2, \hspace{1cm}\boldsymbol{j}=(j_2,\ldots, j_{d_1}) \in \begin{cases}
 \mathbb{N}^{d_1-1}, \;\;\;\;\text{if}\;\; * = D,\\
\tilde{\mathbb{N}}^{d_1-1},\;\;\;\;\text{if}\;\; * =N,\end{cases}
\end{equation*}and the spectrum $\omega_{\mathcal{N};l}^2$ of the Laplace operator on $T^n$ is given by
\begin{equation*}
\left(\frac{l_1}{r_1}\right)^2+\ldots +\left(\frac{l_n}{r_n}\right)^2, \hspace{1cm} \boldsymbol{l} =(l_1, \ldots, l_n) \in \Z^n.
\end{equation*}For simplicity, we only consider the cases where $\alpha=0$ or $1$, i.e., all the walls of the rectangular cavity $[0, L_1]\times\ldots\times[0, L_{d_1}]$ assume Dirichlet boundary conditions or all the walls assume Neumann boundary conditions. The results for the case where $\alpha=1/2$ can be obtained from the results for $\alpha=0$ or $1$ by \eqref{eq3_6_1}. For the finite temperature Casimir energy inside the cavity $[0, L_1]\times [0, L_2]\times\ldots\times[0, L_{d_1}]\times T^n$ given by \eqref{eq2_16_4}, we have
\begin{equation}\label{eq3_3_3}
\begin{split}c_{\text{cyl}, D/N; i}(m) =\sum_{j=0}^{\left[\frac{i}{2}\right]}\frac{(-1)^j}{j!}m^{2j}c_{\text{cyl}, D/N; i-2j},
\end{split}
\end{equation}where
\begin{equation*}
\begin{split}
c_{\text{cyl},   D/N; i}= \frac{(\mp 1)^i }{2^{d_1}\pi^{\frac{d_1-i-n}{2}}}\left[\prod_{l=1}^n r_l\right]    S_{d_1-i}, \hspace{1cm} 0\leq i \leq d_1,
\end{split}
\end{equation*}and $$S_j =\sum_{1\leq \sigma_1<\ldots <\sigma_j\leq d_1}L_{\sigma_1}\ldots L_{\sigma_j}$$is the (hyper)-surface area of $j$-dimensional hyperplanes.  For $i\geq d_1+1$, $c_{\text{cyl}, D/N; i}=0$. The finite temperature zeta function $\zeta_{\text{cyl}, D/N; T}(m)$ can be written as a sum of inhomogeneous Epstein zeta function, i.e.,
\begin{equation*}
\begin{split}
\zeta_{\text{cyl},   D/N; T}(s;m)= 2^{-d_1}\sum_{i=0}^{d_1} (\mp 1)^{d_1-i} \sum_{1\leq \sigma_1<\ldots<\sigma_{i}\leq d_1} Z_{i+n+1}\left(s; \frac{\pi}{L_{\sigma_1}},\ldots,\frac{\pi}{L_{\sigma_{i}}}, \frac{1}{r_1}, \ldots, \frac{1}{r_n}, 2\pi T;m\right),
\end{split}
\end{equation*}where
\begin{equation*}
\begin{split}Z_j \left(s; a_1, \ldots, a_j;m\right)=\sum_{(k_1, \ldots, k_j)\in\Z^j} \frac{1}{\left([k_1a_1]^2+\ldots+[k_ja_j]^2+m^2\right)^s}.
\end{split}
\end{equation*}The properties of the inhomogeneous Epstein zeta function $Z_j \left(s; c_1, \ldots, c_j;m\right)$ has been discussed quite extensively (see e.g. \cite{3_3_1, 3_3_2, 3_3_3, 3_3_4}). In particular, we obtain from  \cite{2_24_1} that if $j$ is even,
\begin{equation}\label{eq3_3_1}
\begin{split}Z_j(0; a_1,\ldots, a_j; m) =&\frac{(-1)^{\frac{j}{2}}}{\left(\frac{j}{2}\right)!}\frac{\pi^{\frac{j}{2}}m^j}{\left[\prod_{i=1}^j a_i\right]},\\
Z_j'(0; a_1,\ldots, a_j; m) =&\frac{(-1)^{\frac{j}{2}}}{\left(\frac{j}{2}\right)!}\frac{\pi^{\frac{j}{2}}m^j}{\left[\prod_{i=1}^j a_i\right]}\left(-\log m^2 +\psi\left(\frac{j+2}{2}\right)-\psi(1)\right)\\&+\frac{2m^{\frac{j}{2}}}{\left[\prod_{i=1}^j a_i\right]}\sum_{(k_1, \ldots, k_j)\in \Z^j\setminus\{0\}}\left(\sum_{i=1}^j \left[\frac{k_i}{a_i}\right]^2\right)^{-\frac{j}{4}}K_{\frac{j}{2}}\left(2\pi m \sqrt{\sum_{i=1}^j \left[\frac{k_i}{a_i}\right]^2}\right);
\end{split}
\end{equation}and if $j$ is odd,
\begin{equation}\label{eq3_3_2}
\begin{split}Z_j(0; a_1,\ldots, a_j; m) =&0,\\
Z_j'(0; a_1,\ldots, a_j; m) =& \frac{\pi^{\frac{j}{2}}m^j}{\left[\prod_{i=1}^j a_i\right]}\Gamma\left(-\frac{j}{2}\right)+\frac{2m^{\frac{j}{2}}}{\left[\prod_{i=1}^j c_i\right]}\sum_{(k_1, \ldots, k_j)\in \Z^j\setminus\{0\}}\left(\sum_{i=1}^j \left[\frac{k_i}{c_i}\right]^2\right)^{-\frac{j}{4}}K_{\frac{j}{2}}\left(2\pi m \sqrt{\sum_{i=1}^j \left[\frac{k_i}{c_i}\right]^2}\right).
\end{split}
\end{equation}Notice that the heat kernel coefficients \eqref{eq3_3_3}   are non-vanishing when $m\rightarrow \infty$. Therefore, they are subtracted away to obtain a regularized Casimir energy that vanishes when $m\rightarrow \infty$. On the other hand, we obtain from \eqref{eq3_3_1} and \eqref{eq3_3_2} that
\begin{equation*}
\begin{split}
-\frac{T}{2}\left( \zeta_{\text{cyl},   D/N; T}'(0;m)+ \log[\mu^2] \zeta_{\text{cyl},   D/N; T}(0;m) \right)= \mathfrak{P} + E_{\text{Cas};D/N}^{\text{reg}}(L_1),
\end{split}
\end{equation*}where $\mathfrak{P}$ is the large-$m$ non-vanishing term and $E_{\text{Cas};D/N}^{\text{reg}}(L_1)$
is  the regularized Casimir energy  given by
\begin{equation}\label{eq3_4_4}
\begin{split}
&E_{\text{Cas};D/N}^{\text{reg}}(L_1) =-\frac{\left[\prod_{l=1}^n r_l\right]}{2^{d_1+1}}\sum_{i=0}^{d_1} \frac{(\mp 1)^{d_1-i}}{\pi^{\frac{i+1-n}{2}}}  m^{\frac{i+n+1}{2}}\sum_{1\leq \sigma_1<\ldots<\sigma_{i}\leq d_1}\left[\prod_{j=1}^i L_{\sigma_j}\right]\sum_{(k_{\sigma_1}, \ldots, k_{\sigma_i}, l_1, \ldots, l_n, p) \in \Z^{i+n+1}\setminus\{0\}}\\
& \times\left(\sum_{j=1}^i \left[L_{\sigma_j}k_{\sigma_j}\right]^2+\sum_{j=1}^n \left[\pi r_j l_j\right]^2+\left[\frac{p}{2T}\right]^2\right)^{-\frac{i+n+1}{4}}K_{\frac{i+n+1}{2}}\left( 2  m \sqrt{\sum_{j=1}^i \left[L_{\sigma_j}k_{\sigma_j}\right]^2+\sum_{j=1}^n \left[\pi r_j l_j\right]^2+\left[\frac{p}{2T}\right]^2}\right).
\end{split}
\end{equation}
For massive scalar field with Neumann boundary conditions, the regularized Casimir energy is negative.

The terms with $p=0$ in \eqref{eq3_4_4} give the zero temperature Casimir energy. The sum of the terms with $\sigma_1\geq 2$ corresponds to the term $\tilde{\Sigma}_0$ in \eqref{eq2_16_10} which is independent of $L_1$. The sum of the terms with $\sigma_1=1$ and $k_{\sigma_1}=k_1=0$ corresponds to the term $\tilde{\Sigma}_1$ in \eqref{eq2_16_10} which is proportional to $L_1$. The rest of the terms decay exponentially when $L_1\rightarrow \infty$.
By applying the formula
\begin{equation*}
\begin{split}
\int_0^{\infty} t^{z-1} \exp\left(-\frac{\beta^2}{t}-t\gamma^2\right)dt =2\left(\frac{\beta}{\gamma}\right)^zK_z(2\beta\gamma)
\end{split}
\end{equation*}and the formula
\begin{equation}\label{eq3_4_1}
\sum_{k=-\infty}^{\infty} \exp\left(-t k^2 \right) = \sqrt{\pi}t^{-\frac{1}{2}} \sum_{k=-\infty}^{\infty} \exp\left(-\frac{\pi^2 k^2}{t}\right),
\end{equation}we can rewrite the regularized Casimir energy $E_{\text{Cas}}^{\text{reg}}(L)$ \eqref{eq3_4_4} in the compact form \eqref{eq2_16_10}.

\section{The Casimir force}\subsection{The Casimir force acting on a piston embedded in a closed cylinder or on two parallel plates embedded in an infinitely long cylinder} \begin{figure}
\epsfxsize=0.5\linewidth \epsffile{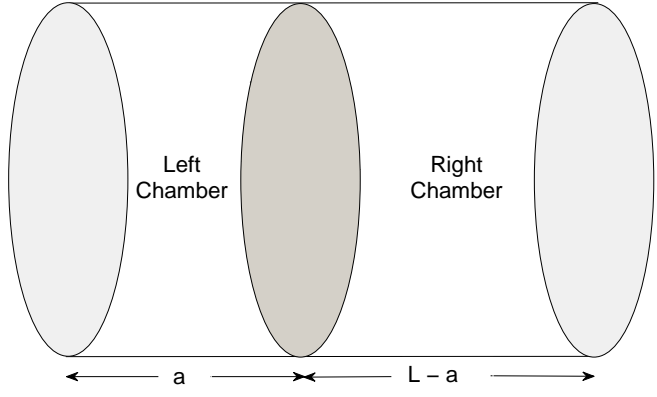} \caption{\label{f1} A movable piston inside a closed cylinder divides the cylinder into two chambers.}\end{figure} To find the Casimir force acting on the walls of the cylinder, one need to take into account the Casimir energy of the outside region, which is not easy to compute. However, there is a setup closely related to the cylinder where the Casimir force can be computed. As in \cite{2_16_1}, one consider a freely moving piston (see FIG. \ref{f1}) dividing the cylinder into two chambers: the left chamber $[0, a]\times \Omega\times\mathcal{N}^n$ and the right chamber $[a, L]\times \Omega\times\mathcal{N}^n$. In this case, the Casimir energy of the region outside the cylinder does not have any effect on the piston. The Casimir force acting on the piston is the sum of the Casimir forces arise from the variations of the Casimir energies in the left and right chambers:
\begin{equation}\label{eq2_16_11}
F_{\text{Cas}}(a;L) = -\frac{\pa}{\pa a}\left( E_{\text{Cas}}(a) +E_{\text{Cas}}(L-a)\right).\end{equation} From \eqref{eq2_16_10}, we find that the contribution to the (regularized)  Casimir force from the left chamber is given by
\begin{equation}\label{eq2_18_2_1}\begin{split}
&F_{\text{Cas}}^{\text{reg, left}}(a)=-\tilde{\Sigma}_1-(-1)^{2\alpha}T  \sum_{j\in J_*}\sum_{l=0}^{\infty} \sum_{p=-\infty}^{\infty} \frac{\sqrt{\omega_{\Omega, *,j}^2+\omega_{\mathcal{N}, l}^2+m^2+(2\pi p T)^2}}{\exp\left(2a\sqrt{\omega_{\Omega, *,j}^2+\omega_{\mathcal{N}, l}^2+m^2+(2\pi p T)^2}\right)-(-1)^{2\alpha}},
\end{split}
\end{equation}where $\tilde{\Sigma}_1$ is a term independent of $a$ given by \eqref{eq2_18_1}. The contribution to the (regularized)  Casimir force from the  right chamber is negative the contribution from the left chamber, with $a$ replaced by $L-a$. As a result, the term $\tilde{\Sigma}_1$ cancels out and we find that the Casimir force acting on the piston is given by
\begin{equation}
F_{\text{Cas}}(a;L)=F_{\text{Cas}}^{\infty}(a)-F_{\text{Cas}}^{\infty}(L-a),
\end{equation}where \begin{equation}\label{eq2_16_12}\begin{split}&
F_{\text{Cas}}^{\infty}(a)=-(-1)^{2\alpha}T  \sum_{j\in J_*}\sum_{l=0}^{\infty} \sum_{p=-\infty}^{\infty} \frac{\sqrt{\omega_{\Omega, *,j}^2+\omega_{\mathcal{N}, l}^2+m^2+(2\pi p T)^2}}{\exp\left(2a\sqrt{\omega_{\Omega, *,j}^2+\omega_{\mathcal{N}, l}^2+m^2+(2\pi p T)^2}\right)-(-1)^{2\alpha}}\end{split}\end{equation} can be interpreted as the limit of the Casimir force when the right chamber is infinitely long.

Since the difference between the cut-off dependent Casimir energy \eqref{eq2_16_4} and the regularized Casimir energy \eqref{eq2_16_10} is given by \eqref{eq2_16_5}, which by \eqref{eq2_16_9} is a linear function in $L$, this implies that we will obtain the same result for the Casimir force whether we use the Casimir energy before or after regularization. In other words, the Casimir force acting on the piston is independent of the regularization procedure employed.

It is easy to deduce from \eqref{eq2_16_12} that the Casimir force for $\alpha=1/2$ is related to the Casimir force for $\alpha=0$ or $1$ by
\begin{equation}\label{eq3_6_2}
\begin{split}
F_{\text{Cas}}\left(a; L; \alpha=\frac{1}{2}; *=D/N\right)=2F_{\text{Cas}}\left(2a; 2L;\alpha=1/0, *=D/N\right)-F_{\text{Cas}}\left(a; L;\alpha=1/0, *=D/N\right),
\end{split}
\end{equation}which also follows from \eqref{eq3_6_1}.

The expression \eqref{eq2_16_12} is negative if $\alpha=0$ or $1$ and positive if $\alpha=1/2$. Moreover, its absolute value is a monotonically decreasing function of $a$. Therefore, if the quantum field assumes either Dirichlet or Neumann boundary conditions on both the piston and the walls of the cylinder, the Casimir force acting on the piston is an attractive force tending to pull the piston towards the closer wall.
If the quantum field assumes Dirichlet (resp. Neumann) boundary conditions on the piston and Neumann (resp. Dirichlet) boundary conditions on the walls of the cylinder, then the Casimir force acting on the piston is a repulsive force tending to restore the piston to its equilibrium position $x^1=L/2$. In both cases, the magnitude of the Casimir force increases as the piston is moving away from the equilibrium position.

Eq. \eqref{eq2_16_12} can also be interpreted as the Casimir force acting on two parallel plates embedded in an infinitely long cylinder with cross section $\Omega\times\mathcal{N}^n$. It shows that the Casimir force  between the plates is attractive if the   field  assumes the same boundary conditions on the plates and is repulsive if the the field  assumes different boundary conditions on the plates, regardless of the boundary conditions assumed on the surrounding transversal wall. Writing $\omega_{\Omega, *;j}=\omega_{\Omega,*;j}'/R$ and $\omega_{\mathcal{N};l} =\omega_{\mathcal{N};l}'/r$, eq. \eqref{eq2_16_12}   shows that when   $a/r$ or $a/R$ or    $am$ is large,  the magnitude of the Casimir force decays exponentially. At high temperature $T$, eq. \eqref{eq2_16_12} shows that the leading term of the Casimir force is given by a term linear in $T$:
\begin{equation}\label{eq2_17_1}\begin{split}&
F_{\text{Cas}}^{\infty, T\gg 1}(a)\sim -(-1)^{2\alpha}T  \sum_{j\in J_*}\sum_{l=0}^{\infty}   \frac{\sqrt{\omega_{\Omega, *,j}^2+\omega_{\mathcal{N}, l}^2+m^2 }}{\exp\left(2a\sqrt{\omega_{\Omega, *,j}^2+\omega_{\mathcal{N}, l}^2+m^2 }\right)-(-1)^{2\alpha}},\end{split}\end{equation}and the remaining term decays exponentially.
Notice that if we consider the contribution to the Casimir force from the left chamber \eqref{eq2_18_2_1}, then the result \eqref{eq2_18_2} of Appendix \ref{a1} shows that when $T\gg 1$,
\begin{equation}\label{eq2_18_3}
\begin{split}
F_{\text{Cas}}^{\text{reg, left}}(a)\sim &\frac{1}{4\pi}\sum_{i=0}^{d-1} \Gamma\left(\frac{i+2}{2}\right)\zeta_R(i+2)  c_{\Omega\times\mathcal{N},*;d-1-i}(m)(2T)^{i+2}+O(T).
\end{split}
\end{equation}We observe that in general, there are terms of order $T^{2}, T^3, \ldots, T^{d+1}$. In particular, the leading term is
\begin{equation*}\begin{split}
F_{\text{Cas}}^{\text{reg, left}}(a)\sim &\frac{2^{d-1}\Gamma\left(\frac{d+1}{2}\right)\zeta_R(d+1)}{\pi} c_{\Omega\times\mathcal{N},*;0}(m)T^{d+1}
=\frac{ \Gamma\left(\frac{d+1}{2}\right)\zeta_R(d+1)}{\pi^{\frac{d+1}{2}}} \text{Vol}(\Omega\times\mathcal{N})T^{d+1},
\end{split}\end{equation*}which is the Stefan-Boltzmann term. Usually this term is subtracted away in the regularization of the Casimir energy since it can be interpreted as the contribution to the vacuum energy in the absence of boundaries.  However, besides the limiting case of infinite parallel plates, there are still terms with order $T^2, \ldots, T^d$. Nevertheless, since  these terms   are independent of $a$, they cancel   the corresponding contributions from the right chamber and the high temperature leading term of the  Casimir force acting on the piston is a term linear in $T$ given by \eqref{eq2_17_1}. This is usually called the classical term \cite{1_15_1, 1_15_2, 1_15_3, 1_15_4} due to the absence of the Planck constant $\hbar$ in this term. In a recent work on  Casimir effect of electromagnetic field in three dimensional ideal metal rectangular box \cite{nn1}, it has been argued that the terms of order $T^2, \ldots, T^{d+1}$ have to be subtracted away in order to be consistent with thermodynamics. Here we find that in the piston scenario, such terms are naturally absent due to the cancelation between the two sides of the plate.

For the low temperature asymptotic expansion of the Casimir force,  we use the formula
\begin{equation}\label{eq2_16_13}
\begin{split}
&  \frac{(-1)^{2\alpha-1}\sqrt{\omega_{\Omega, *,j}^2+\omega_{\mathcal{N}, l}^2+m^2+(2\pi p T)^2}}{\exp\left(2a\sqrt{\omega_{\Omega, *,j}^2+\omega_{\mathcal{N}, l}^2+m^2+(2\pi p T)^2}\right)-(-1)^{2\alpha}} \\=&\frac{1}{2\sqrt{\pi}}\frac{\pa}{\pa a}\Biggl\{ a\int_0^{\infty} t^{-\frac{1}{2}}\sum_{k=1}^{\infty} e^{2\pi i k\alpha}\exp\left\{-\frac{1}{t}\left(\omega_{\Omega,*;j}^2+\omega_{\mathcal{N}, l}^2+[2\pi p T]^2+m^2\right)-tk^2 a^2\right\}dt\Biggr\}
\end{split}
\end{equation}and the formula \eqref{eq3_4_1}. These give
\begin{equation}\label{eq2_17_8}
\begin{split}
F_{\text{Cas}}^{\infty}(a) = &\frac{1}{4\pi}\frac{\pa}{\pa a}\Biggl\{ a\int_0^{\infty}  \sum_{k=1}^{\infty}\sum_{j\in J_*}\sum_{l=0}^{\infty}\sum_{p=-\infty}^{\infty} e^{2\pi i k\alpha} \exp\left\{-\frac{ \omega_{\Omega,*;j}^2+\omega_{\mathcal{N}, l}^2 +m^2}{t}-t\left([k a]^2 +\left[\frac{p}{2T}\right]^2\right)\right\}dt\Biggr\}\\
=&\frac{1}{2\pi} \sum_{k=1}^{\infty}\sum_{j\in J_*}\sum_{l=0}^{\infty}\sum_{p=-\infty}^{\infty} e^{2\pi i k\alpha}\Biggl\{
\sqrt{\frac{ \omega_{\Omega,*;j}^2+\omega_{\mathcal{N}, l}^2 +m^2}{[k a]^2 +\left[\frac{p}{2T}\right]^2}} K_1\left(2\sqrt{\left(\omega_{\Omega,*;j}^2+\omega_{\mathcal{N}, l}^2 +m^2\right)\left([k a]^2 +\left[\frac{p}{2T}\right]^2\right)}\right)\\
&-2k^2a^2 \frac{ \omega_{\Omega,*;j}^2+\omega_{\mathcal{N}, l}^2 +m^2}{[k a]^2 +\left[\frac{p}{2T}\right]^2} K_2\left(2\sqrt{\left(\omega_{\Omega,*;j}^2+\omega_{\mathcal{N}, l}^2 +m^2\right)\left([k a]^2 +\left[\frac{p}{2T}\right]^2\right)}\right)
\Biggr\}.
\end{split}
\end{equation}
In the zero temperature limit, we obtain the zero temperature Casimir force from the terms with $p=0$:
\begin{equation}\label{eq3_3_4}
\begin{split}
F_{\text{Cas}}^{\infty, T=0}(a) =&\frac{1}{2\pi} \sum_{k=1}^{\infty}\sum_{j\in J_*}\sum_{l=0}^{\infty}  e^{2\pi i k\alpha}\Biggl\{
 \frac{ \sqrt{\omega_{\Omega,*;j}^2+\omega_{\mathcal{N}, l}^2 +m^2}}{k a } K_1\left(2ka\sqrt{ \omega_{\Omega,*;j}^2+\omega_{\mathcal{N}, l}^2 +m^2}\right) \\
&-2 \left(\omega_{\Omega,*;j}^2+\omega_{\mathcal{N}, l}^2 +m^2\right) K_2\left(2ka\sqrt{ \omega_{\Omega,*;j}^2+\omega_{\mathcal{N}, l}^2 +m^2}\right)
\Biggr\},
\end{split}
\end{equation}which can also be derived directly from eq. \eqref{eq2_16_1}. The temperature correction  to the Casimir force is the sum of the terms with $p\neq 0$ in \eqref{eq2_17_8}. It goes to zero exponentially fast when $T\rightarrow 0$. Using the identity $K_2(z) = K_0(z)+2K_1(z)/z$, we can rewrite the zero temperature Casimir force \eqref{eq3_3_4} as
\begin{equation}\label{eq3_3_5}
\begin{split}
F_{\text{Cas}}^{\infty, T=0}(a) =&-\frac{1}{2\pi} \sum_{k=1}^{\infty}\sum_{j\in J_*}\sum_{l=0}^{\infty}  e^{2\pi i k\alpha}\Biggl\{
 \frac{ \sqrt{\omega_{\Omega,*;j}^2+\omega_{\mathcal{N}, l}^2 +m^2}}{k a } K_1\left(2ka\sqrt{ \omega_{\Omega,*;j}^2+\omega_{\mathcal{N}, l}^2 +m^2}\right) \\
&+2 \left(\omega_{\Omega,*;j}^2+\omega_{\mathcal{N}, l}^2 +m^2\right) K_0\left(2ka\sqrt{ \omega_{\Omega,*;j}^2+\omega_{\mathcal{N}, l}^2 +m^2}\right)
\Biggr\},
\end{split}
\end{equation}which shows manifestly that the zero temperature Casimir force is attractive if $\alpha=0$ or $1$ and is repulsive if $\alpha=1/2$.

For the behavior of the Casimir force with respect to the variation of mass,  we observe that the function $$x\mapsto \frac{x}{e^x-1}$$ is a decreasing function. Therefore  the magnitude of the Casimir force acting between a pair of parallel plates embedded in an infinitely long cylinder (of arbitrarcy cross section) is a decreasing function of $m$ when $\alpha=0$ or $1$. In other words, for a pair of parallel plates with identical boundary conditions, the increase in mass   reduces the Casimir effect. This property cannot be obviously inferred from the expressions for the zero temperature Casimir force given by \eqref{eq3_3_4} or \eqref{eq3_3_5}. We see here the advantage of considering the Casimir force at any finite temperature. It enables us to derive some properties of the Casimir force from the expression \eqref{eq2_16_12}, which cannot be derived directly from the expression for zero temperature Casimir force \eqref{eq3_3_4} or \eqref{eq3_3_5}.

For a pair of parallel plates with mixed boundary conditions or when   $\alpha=1/2$, the situation is different. The function $$x\mapsto \frac{x}{e^x+1}$$ is increasing when $x\in [0, 1.2785]$ and decreasing for $x\geq 1.2785$. Therefore, when the mass increases, the magnitude of the Casimir force may first increase and then decrease exponentially.

Notice that when the mass $m$ decreases to zero, \eqref{eq2_16_12} naively shows that the Casimir force tends to the Casimir force for massless scalar fields \cite{2_10_1}. To be more careful, we need to discuss the cases where the field assumes Dirichlet boundary conditions and Neumann boundary conditions on the surrounding walls $[0,L]\times \pa\Omega\times\mathcal{N}^n$ separately. If the field assumes Dirichlet boundary conditions on the wall $[0,L]\times \pa\Omega\times\mathcal{N}^n$, then the Dirichlet eigenvalues $\omega_{\Omega, D;j}^2$ are all nonzero. In this case, we can immediately set $m=0$ in \eqref{eq2_16_12} and obtain the Casimir force for massless scalar field. However, if the field assumes Neumann boundary conditions on the wall $[0,L]\times \pa\Omega\times\mathcal{N}^n$, there is exactly one zero Neumann eigenvalue $\omega_{\Omega,N;0}^2$. In this case,  there are $\kappa$ pairs of $(j,l)$ such that $\omega_{\Omega,N;j}^2+\omega_{\mathcal{N};l}^2=0$. Separating the sum over $(j,l)\in J_N \times \tilde{\mathbb{N}}$ in \eqref{eq2_16_12} as a sum over those $(j,l)$ with $\omega_{\Omega,N;j}^2+\omega_{\mathcal{N};l}^2=0$ and those $(j,l)$ with $\omega_{\Omega,N;j}^2+\omega_{\mathcal{N};l}^2\neq 0$, we find that the we can immediately set $m=0$ in the latter sum to obtain the contribution of the modes with $\omega_{\Omega,N;j}^2+\omega_{\mathcal{N};l}^2\neq 0$ to the Casimir force due to massless scalar field. For the sum over   $(j,l)$ with $\omega_{\Omega,N;j}^2+\omega_{\mathcal{N};l}^2=0$ given by
\begin{equation}\label{eq2_17_2}
\begin{split}
-(-1)^{2\alpha}\kappa T\sum_{p=-\infty}^{\infty} \frac{\sqrt{m^2+(2\pi pT)^2}}{\exp\left(2a\sqrt{m^2+(2\pi pT)^2}\right)-(-1)^{2\alpha}},
\end{split}
\end{equation}we have to be careful when taking the massless limit for the term with $p=0$. Using the fact that $$\frac{x}{e^x-1}=1 + O(x), \hspace{1cm} \frac{x}{e^x+1}=O(x)\hspace{1cm}\text{as}\;x\rightarrow 0,$$we find that the massless limit of \eqref{eq2_17_2} is given by
\begin{equation}\label{eq2_17_3}
\kappa \left\{-|2\alpha-1|\frac{T}{2a} +4\pi T^2\sum_{p=1}^{\infty}\frac{(-1)^{2\alpha-1}p}{\exp(4\pi p Ta)-(-1)^{2\alpha}}\right\}.
\end{equation}Except for the factor $\kappa$, this is the finite temperature Casimir force between a pair of parallel plates in (1+1)-dimensional Minkowski spacetime due to massless scalar field with Dirichlet or Neumann boundary conditions on both plates (for $\alpha=0$ or $1$) and with Dirichlet boundary condition on one plate and Neumann boundary condition on the other plate (for $\alpha=1/2$) (see \cite{2_17_1}). As in \cite{2_17_1}, the Casimir force due to the modes with $\omega_{\Omega,N;j}^2+\omega_{\mathcal{N};l}^2=0$ \eqref{eq2_17_3} has an alternative expression given by
\begin{equation}\label{eq2_17_4}
\kappa\left(-\frac{\pi}{24a^2}-\frac{\pi T^2} {6} +\frac{\pi}{a^2}\sum_{k=1}^{\infty}\frac{k}{\exp\left(\frac{\pi k}{Ta}\right)-1}\right), \;\;\;\;\text{if}\;\alpha=0,1;
\end{equation}
and \begin{equation}\label{eq2_17_5}
\kappa\left(\frac{\pi}{48a^2}-\frac{\pi T^2} {6} +\frac{\pi}{a^2}\sum_{k=1}^{\infty}\frac{k+\frac{1}{2}}{\exp\left(\frac{\pi \left(k+\frac{1}{2}\right)}{Ta}\right)-1}\right), \;\;\;\;\text{if}\;\alpha=\frac{1}{2}.
\end{equation}The first term in \eqref{eq2_17_4} and \eqref{eq2_17_5} give the respective zero temperature Casimir force. Notice that in contrast to the massive case where the force decays exponentially, they decay in the order $1/a^2$ when $a$ is large. This is a long range force which is the subject of study in the context of electromagnetic fields in the recent work \cite{2_17_2}. At finite temperature, \eqref{eq2_17_3} shows that this long range force is present if and only if $\alpha\neq 1/2$, i.e., if and only if the boundary conditions assumed on the  two plates and the walls of the cylinder are both Neumann  conditions. In this case, the long range force is of order $T/a$. From this analysis, we find that long range force may exist only in the massless case when Neumann boundary conditions are assumed on  the wall of the cylinder and the two plates or when the temperature is zero and Dirichlet boundary conditions are assumed on the two plates. The  transition  from massless to massive field will change the nature of the force from long range to short range.

To investigate the dependence of the Casimir force $F_{\text{Cas}}^{\infty}(a)$ on the size $R$ of the cross section $\Omega$,  we rewrite $\omega_{\Omega, *; j}$ as $\omega_{\Omega, *; j}'/R$, where the re-scaled frequency $\omega_{\Omega,*; j}'$ is independent of the size $R$ of the cross section $\Omega$. Using the same argument about the dependence of the Casimir force on mass, we see that if $\alpha=0$ or $1$, the Casimir force increases when the size $R$  increases. The asymptotic behaviors of the Casimir force when the plate separation $a$ is much smaller than the size $R$ of the cross section $\Omega$ is derived in Appendix \ref{a2}. We read from \eqref{eq2_23_1} and \eqref{eq2_23_3} that if $a\ll R$, then if $aT\gg 1$,
\begin{equation}\label{eq2_23_2}
\begin{split}
F^{\infty}_{\text{Cas}}(a) \sim &\frac{T}{a\sqrt{\pi}}\sum_{i=0}^{d_1-1}c_{\Omega/R,*;i} \left(\frac{R}{a}\right)^{d_1-i-1}\sum_{k=1}^{\infty}\sum_{l=0}^{\infty}\sum_{p=-\infty}^{\infty}e^{2\pi i k\alpha}\Biggl\{ \left(\frac{a\sqrt{\omega_{\mathcal{N};l}^2+(2\pi p T)^2+m^2}}{k}\right)^{\frac{d_1 -i}{2}}\\&\times K_{\frac{d_1-i }{2}}\left(2ka \sqrt{\omega_{\mathcal{N};l}^2+(2\pi p T)^2+m^2}\right)-2\frac{\left(a\sqrt{\omega_{\mathcal{N};l}^2+(2\pi p T)^2+m^2}\right)^{\frac{d_1+2-i }{2}}}{k^{\frac{d_1-2-i }{2}}}\\&\times K_{\frac{d_1+2-i }{2}}\left(2ka \sqrt{\omega_{\mathcal{N};l}^2+(2\pi p T)^2+m^2}\right)\Biggr\}+O\left( \left(\frac{R}{a}\right)^{-1}\right),
\end{split}
\end{equation}and if $aT\ll 1$,
\begin{equation}\label{eq2_23_4}\begin{split}F^{\infty}_{\text{Cas}}(a)\sim & -\frac{1}{a^2}\sum_{i=0}^{d_1-1} c_{\Omega/R, *; i}\left(\frac{R}{a}\right)^{d_1-1-i} \Biggl\{ \frac{1}{2\pi}\sum_{k=1}^{\infty}e^{2\pi i k\alpha} \sum_{l=0}^{\infty}\Biggl[ (d_1-i) \left( \frac{a\sqrt{\omega_{\mathcal{N};l}^2+m^2}}{k}\right)^{\frac{d_1+1-i}{2}}K_{\frac{d_1+1-i}{2}}\left( 2ka\sqrt{\omega_{\mathcal{N};l}^2+m^2}\right)\\&+2
\frac{\left(a\sqrt{\omega_{\mathcal{N};l}^2+m^2}\right)^{\frac{d_1+3-i}{2}}}{k^{\frac{d_1-1-i}{2}}}K_{\frac{d_1-1-i}{2}}\left( 2ka\sqrt{\omega_{\mathcal{N};l}^2+m^2}\right)\Biggr]- 2^{\frac{d_1-i}{2}}\pi^{\frac{3}{2}}(aT)^{\frac{d_1-i-2}{2}} \sum_{k=0}^{\infty} \sum_{l=0}^{\infty}\sum_{p=1}^{\infty} (k+\alpha)^2 \\&\times \left(\frac{ \sqrt{[\pi (k+\alpha)]^2+(a\omega_{\mathcal{N};l})^2+(am)^2}}{p}\right)^{\frac{d_1-i-2}{2}}K_{\frac{d_1-i-2}{2}}\left(\frac{p}{Ta }\sqrt{[\pi ( k+\alpha)]^2+(a\omega_{\mathcal{N};l})^2+(am)^2}\right)\\
&+\frac{(2aT)^{\frac{d_1-i+1}{2}}}{2\pi}\sum_{l=0}^{\infty}\sum_{p=1}^{\infty}\left(\frac{ a\sqrt{\omega_{\mathcal{N};l}^2+m^2}}{p}\right)^{\frac{d_1-i+1}{2}}K_{\frac{d_1-i+1}{2}}\left(\frac{p}{T}\sqrt{\omega_{\mathcal{N};l}^2+m^2}\right)\Biggr\}+O\left( \left(\frac{R}{a}\right)^{-1}\right).
\end{split}
\end{equation}Here $c_{\Omega/R,*;i}$ are the heat kernel coefficients of the Laplace operator with Dirichlet ($*=D$) or Neumann ($*=N$) boundary conditions on $\Omega/R$. Notice that when $r\ll a\ll R$, the   large--$R$ non-vanishing terms of the Casimir force can be written as a polynomial of order $d_1-1$ in $R$ with coefficients depending on the geometric invariants $c_{\Omega, *;i}$ of $\Omega$, and   Bessel series that depend on the geometry of the internal manifold $\mathcal{N}^n$, and the plate separation $a$. From these expressions, it is easy to read that if the   mass $m$ is also very large, the Casimir force decays  exponentially. In the case   the mass $m$ is small, or more precisely if $am\ll 1\ll Rm$, we obtain from \eqref{eq2_23_5} and \eqref{eq2_26_1} that if $\alpha=0$ or $1$,
\begin{equation*}
\begin{split}& F^{\infty}_{\text{Cas}}(a)\sim  \frac{\kappa T}{2\sqrt{\pi}a}\sum_{j=1}^{\left[\frac{d_1}{2}\right]}c_{\Omega/R,*;d_1-2j}\left(\frac{R } {a}\right)^{2j-1} \Biggl\{\frac{1}{2}\frac{(-1)^{j}}{j!}(am)^{2j}\left(\log\left(\frac{am}{2\pi}\right)^2+2-\psi\left(j+1\right)-\psi(1)\right)\\& + \sqrt{\pi} \sum_{q=0}^{j-1}\frac{(-1)^q}{q!}(am)^{2q}  (2q-2j+1)\pi^{2j-2q-1} \Gamma\left(-j+\frac{1}{2}+q\right) \zeta_R(-2j+1+2q) \Biggr\}  + \frac{\kappa T}{2\sqrt{\pi}a}\sum_{j=0}^{\left[\frac{d_1-1}{2}\right]} c_{\Omega/R,*;d_1-2j-1}  \left(\frac{R}{a}\right)^{2j} \\&\times \Biggl\{ -\sqrt{\pi}\frac{(-1)^{j}}{j!}(am)^{2j} -\frac{1}{2}\Gamma\left(-j-\frac{1}{2}\right)(am)^{2j+1}   + \sqrt{\pi} \sum_{q=0}^{j-1}\frac{(-1)^q}{q!}(am)^{2q}(2q-2j)\pi^{2j-2q} \Gamma\left(-j+q\right) \zeta_R(-2j+2q) \Biggr\}\end{split}\end{equation*}\begin{equation}\label{eq2_24_1}\begin{split}&+\frac{T}{\sqrt{\pi} a}\sum_{i=0}^{d_1-1} c_{\Omega,*;i}\left(\frac{R}{ a}\right)^{d_1-1-i}\sum_{q=0}^{d_1+1-i}\frac{(-1)^q}{q!}(am)^{2q}\sum_{k=1}^{\infty}\sum_{\substack{(l,p)\in \tilde{\mathbb{N}}\times\Z\\
\omega_{\mathcal{N};l}^2+(2\pi pT)^2\neq 0}}\Biggl\{ \left(\frac{a\sqrt{\omega_{\mathcal{N};l}^2+(2\pi p T)^2}}{k}\right)^{\frac{d_1-i-2q}{2}}\\&\times K_{\frac{d_1-i-2q}{2}}\left(2ka \sqrt{\omega_{\mathcal{N};l}^2+(2\pi p T)^2}\right)-2\frac{\left(a\sqrt{\omega_{\mathcal{N};l}^2+(2\pi p T)^2}\right)^{\frac{d_1+2-i-2q}{2}}}{k^{\frac{d_1-2-i-2q}{2}}}K_{\frac{d_1+2-i-2q}{2}}\left(2ka \sqrt{\omega_{\mathcal{N};l}^2+(2\pi p T)^2}\right)\Biggr\},
\end{split}
\end{equation}if $aT\gg 1$; and
\begin{equation}\label{eq2_24_2}
\begin{split}& F^{\infty}_{\text{Cas}}(a)\sim \frac{ \kappa}{4\pi a^2}\sum_{j=1}^{\left[\frac{d_1+1}{2}\right]}c_{\Omega/R,*;d_1+1-2j}\left(\frac{R } {a}\right)^{2j-2} \Biggl\{\frac{1}{2}\frac{(-1)^{j}}{j!}(am)^{2j}\left(\log\left(\frac{am}{2\pi}\right)^2+2-\psi\left(j+1\right)-\psi(1)\right) \\&+ \sqrt{\pi} \sum_{q=0}^{j-1}\frac{(-1)^q}{q!}(am)^{2q} (2q-2j+1)\pi^{2j-2q-1} \Gamma\left(-j+\frac{1}{2}+q\right) \zeta_R(-2j+1+2q) \Biggr\}  +\frac{\kappa }{4\pi a^2}\sum_{j=0}^{\left[\frac{d_1}{2}\right]} c_{\Omega/R,*;d_1-2j} \left(\frac{R}{a}\right)^{2j-1}\\&\times \Biggl\{ -\sqrt{\pi}\frac{(-1)^{j}}{j!}(am)^{2j} -\frac{1}{2}\Gamma\left(-j-\frac{1}{2}\right)(am)^{2j+1}   + \sqrt{\pi} \sum_{q=0}^{j-1}\frac{(-1)^q}{q!}(am)^{2q}(2q-2j)\pi^{2j-2q} \Gamma\left(-j+q\right) \zeta_R(-2j+2q) \Biggr\}\\&-\frac{1}{2\pi a^2}\sum_{ i=0}^{d_1-1} c_{\Omega,*;i}\left(\frac{R}{ a}\right)^{d_1-1-i}\sum_{q=0}^{d_1+1-i}\frac{(-1)^q}{q!}(am)^{2q}\sum_{k=1}^{\infty}\sum_{\substack{l\in \tilde{\mathbb{N}}\\ \omega_{\mathcal{N},l}^2\neq 0}}\Biggl\{ (d_1-i-2q) \left(\frac{\omega_{\mathcal{N};l}}{k}\right)^{\frac{d_1+1-i-2q}{2}}K_{\frac{d_1+1-i-2q}{2}}\left(2ka\omega_{\mathcal{N};l}\right)\\
&+2\frac{\omega_{\mathcal{N};l}^{\frac{d_1+3-i-2q}{2}}}{k^{\frac{d_1-1-i-2q}{2}}}K_{\frac{d_1-1-i-2q}{2}}\left(2ka\omega_{\mathcal{N};l}\right)\Biggr\}
\end{split}
\end{equation}
if $T=0$. The omitted terms goes to zero when $am\rightarrow 0$ or $R/a\rightarrow \infty$. Notice that these asymptotic behaviors contain logarithmic terms in $am$ which goes to zero when the mass $m$ approaches zero. In the massless case, we have
\begin{equation*}
\begin{split}F_{\text{Cas}}^{\infty}(a; m=0) \sim -\frac{\kappa T}{2\sqrt{\pi} }\sum_{i=0}^{d_1-2} \frac{c_{\Omega,*; i} }{a^{d_1-i}}(d_1-i-1) \Gamma\left(\frac{d_1-i}{2}\right)\zeta_R(d_1-i)-\frac{\kappa T}{2}\frac{c_{\Omega, *; d_1-1}}{a} +O(a^0), \hspace{1cm}\text{as} \;\; r\ll a\rightarrow 0^+
\end{split}
\end{equation*}if $aT\gg 1$; and
\begin{equation*}
F_{\text{Cas}}^{\infty}(a; m=0) \sim -\frac{\kappa}{4\pi }\sum_{i=0}^{d_1-1}\frac{c_{\Omega, *; i}}{a^{d_1-i+1}}(d_1-i)\Gamma\left(\frac{d_1-i+1}{2}\right) \zeta_R(d_1-i+1) -\frac{\kappa}{4\sqrt{\pi}}\frac{c_{\Omega, *; d_1}}{a} +O(a^0), \hspace{1cm}\text{as} \;\; r\ll a\rightarrow 0^+
\end{equation*}if $aT\ll 1$. From \eqref{eq2_24_1} and \eqref{eq2_24_2}, we find that when $m$ is small, then as $a\rightarrow 0^+$, the leading order term of the Casimir force is given by
\begin{equation*}
F_{\text{Cas}}^{\infty}(a) \sim -(d_1-1)\Gamma\left(\frac{d_1}{2}\right)\zeta_R(d_1)\frac{\text{Vol}(\Omega)}{(4\pi)^{\frac{d_1}{2}}}\frac{\kappa T}{a^{d_1+1}}, \hspace{1cm}\text{if}\;\;aT\gg 1,
\end{equation*}and
\begin{equation*}
F_{\text{Cas}}^{\infty}(a) \sim -d_1\Gamma\left(\frac{d_1+1}{2}\right)\zeta_R(d_1+1)\frac{\text{Vol}(\Omega)}{(4\pi)^{\frac{d_1+1}{2}}}\frac{\kappa }{a^{d_1}}, \hspace{1cm}\text{if}\;\;aT\ll 1,
\end{equation*}respectively, which are independent of the mass $m$. This shows that when $am\ll 1$, the mass correction  to the Casimir force is not significant.

If $d_1=3$ and $\alpha=0$ or $1$, \eqref{eq2_24_1} and \eqref{eq2_24_2} give respectively
\begin{equation}\label{eq2_24_3}
\begin{split}
F^{\infty}_{\text{Cas}}(a)\sim &\kappa T\left\{-\frac{c_{\Omega/R,*;0}R^2}{2a^3}\zeta_R(3) -\frac{\pi^{\frac{3}{2}}c_{\Omega/R,*;1}R}{12a^2}-\frac{c_{\Omega/R,*;2}}{2a} +\frac{m^2c_{\Omega/R,*; 0 }R^2}{2a}-\frac{m^2 c_{\Omega/R,*;1} R}{4\sqrt{\pi } }\log(am)^2 \right\}+O(a^0)\\&\text{as}\;\; r\ll a\rightarrow 0^+
\end{split}
\end{equation}if $T\gg 1$, and
\begin{equation}\label{eq2_24_4}
\begin{split}
F^{\infty}_{\text{Cas}}(a)\sim  \kappa\Biggl\{ &-\frac{\pi^3 c_{\Omega/R,*;0} R^2}{120a^4}-\frac{c_{\Omega/R,*;1}R}{4\sqrt{\pi} a^3}\zeta_R(3)-\frac{\pi c_{\Omega/R, *;2}}{24a^2}
+\frac{m^2 c_{\Omega/R,*;1}R}{4\sqrt{\pi} a}\\&-\frac{c_{\Omega/R, *;2}}{8\pi} m^2\log(am)^2+\frac{m^4c_{\Omega/R,*;0}R^2}{16\pi}\log(am)^2\Biggr\}+O(a^0)\hspace{1cm}\text{as}\;\; r\ll a\rightarrow 0^+,
\end{split}
\end{equation}if $T\ll 1$. If $\alpha=1/2$, similar computation gives
 \begin{equation*}
\begin{split}
F^{\infty}_{\text{Cas}}(a)\sim &\kappa T\left\{\frac{3 c_{\Omega/R,*;0}R^2}{8a^3}\zeta_R(3) +\frac{\pi^{\frac{3}{2}}c_{\Omega/R,*;1}R}{24a^2} -\frac{m^2 c_{\Omega/R,*;1} R }{4\sqrt{\pi } }\log (am)^2\right\}+O(a^0)\hspace{1cm}\text{as}\;\; r\ll a\rightarrow 0^+
\end{split}
\end{equation*}if $T\gg 1$, and
\begin{equation*}
\begin{split}
F^{\infty}_{\text{Cas}}(a)\sim  \kappa\Biggl\{ &\frac{7\pi^3 c_{\Omega/R,*;0} R^2}{960a^4}+\frac{3c_{\Omega/R,*;1}R}{16\sqrt{\pi} a^3}\zeta_R(3)+\frac{\pi c_{\Omega/R, *;2}}{48a^2}
 \\&-\frac{c_{\Omega/R, *;2}}{8\pi} m^2\log(am)^2+\frac{m^4c_{\Omega/R,*;0}}{16\pi}\log(am)^2\Biggr\}+O(a^0)\hspace{1cm}\text{as}\;\; r\ll a\rightarrow 0^+,
\end{split}
\end{equation*}if $T\ll 1$. When $d_1=3$ and $\Omega=[0, L_2]\times [0, L_3]$ is a rectangle, \eqref{eq2_24_4} gives the correct behavior of the Casimir force when $a \ll 1$ which was derived in \cite{2_24_1}.

In \eqref{eq2_24_2}, we only give the asymptotic behavior of the zero temperature Casimir force when $am\ll 1\ll Rm$. At finite low temperature, the behavior of the thermal correction depends on the relative strength of the temperature $T$ and mass $m$. If $T\ll m$, the thermal correction term can be expanded in a power series of $m^2$ plus an   exponentially suppressed term. If $T\gg m$, then the thermal correction term can be written as a sum of a power series in $m^2$ as in the case of $T\ll m$ and the term \eqref{eq3_2_1}, which  contains logarithmic terms of $m/T$.

For the influence of the extra dimensions,    we rewrite $\omega_{\mathcal{N}; l}$ as $\omega_{\mathcal{N}; l}'/r$, where the re-scaled frequency $\omega_{\mathcal{N}; l}'$ is independent of the size $r$ of the extra dimensions $\mathcal{N}^n$. We find from each of the formulas and asymptotic expansions for the Casimir force $F_{\text{Cas}}^{\infty}(a)$ derived above that, as the size $r$ is very small compared to the plate separation $a$, the terms with nonzero $\omega_{\mathcal{N};l}^2$ contribute Casimir force that are exponentially small. In the limit the internal manifold $\mathcal{N}^n$ vanishes, i.e. $r\rightarrow 0^+$, only the $\kappa$ terms corresponding to $\omega_{\mathcal{N};l}^2=0$ remain and they give $\kappa$ times the Casimir force in $(d_1+1)$-dimensional Minkowski spacetime. In other words, if $\kappa=0$, the Casimir force goes to zero in the limit of vanishing internal space. This is definitely not a desired physical situation. The situation  that is of physical interest is the recovery of the  Casimir force in $(d_1+1)$-dimensional Minskowski spacetime in the limit of vanishing extra dimensions, or equivalently $\kappa=1$.  This happens in particular when $\mathcal{N}^n$ has zero scalar curvature or when $\xi=0$ (minimal coupling) and $\mathcal{N}^n$ is connected. In this case, \eqref{eq2_16_12} shows that the presence of extra dimensions enhances the Casimir effect. In case $\alpha=0$ or $1$, the expression \eqref{eq2_16_12} shows that the magnitude of the Casimir force becomes larger when the size of the internal manifold $r$ is larger.

  When the size $r$ of the internal manifold $\mathcal{N}^n$ becomes comparable to the plate separation $a$, the correction to the Casimir force in $(d_1+1)$-dimensional Minskowki spacetime due to the extra dimensions (i.e. the contributions from the terms with $\omega_{\mathcal{N}; l}^2\neq 0$) can become substantial and it depends on the geometry of the internal manifold $\mathcal{N}^n$.

\subsection{The Casimir force density acting on a pair of parallel plates}

In this section, we consider the limit where the size  $R$ of the cross section $\Omega$ goes to infinity, which is tantamount to two parallel plates in a $(d_1+1)$-dimensional Minkowski  spacetime $M^{d_1+1}$, with an $n$-dimensional internal  manifold compactified to $\mathcal{N}^n$ at every point of $M^{d_1+1}$. In this case, we should consider the Casimir force density $\mathcal{F}_{\text{Cas}}^{\parallel}(a)$ on the plates $x^1=0$ and $x^1=a$ which is defined as the limit
\begin{equation*}
\mathcal{F}_{\text{Cas}}^{\parallel}(a)=\lim_{R\rightarrow \infty} \frac{F_{\text{Cas}}^{\infty}(a)}{\text{Vol}(\Omega)} =\lim_{R\rightarrow \infty}\frac{F_{\text{Cas}}^{\infty}(a)}{R^{d_1-1}}.
\end{equation*}From \eqref{eq2_23_2} and \eqref{eq2_23_4}, we observe that when $R$ is large, the leading term of $F^{\infty}_{\text{Cas}}(a)$ is of order $R^{d_1-1}$ coming from the term with $i=0$. All the remaining terms are of order smaller than $R^{d_1-1}$. Using the fact that $c_{\Omega/R,*;0} =1/(2\sqrt{\pi})^{d_1-1}$, we obtain immediately the following high and low temperature expansions for the Casimir force density $\mathcal{F}_{\text{Cas}}^{\parallel}(a)$:
\begin{equation*}
\begin{split}\mathcal{F}_{\text{Cas}}^{\parallel}(a)=&\frac{1}{2^{d_1-1}\pi^{\frac{d_1}{2}}}\frac{T}{a^{d_1}} \sum_{k=1}^{\infty}\sum_{l=0}^{\infty}\sum_{p=-\infty}^{\infty}e^{2\pi i k\alpha}  \left(\frac{a\sqrt{\omega_{\mathcal{N};l}^2+(2\pi p T)^2+m^2}}{k}\right)^{\frac{d_1}{2}}K_{\frac{d_1 }{2}}\left(2ka \sqrt{\omega_{\mathcal{N};l}^2+(2\pi p T)^2+m^2}\right)\\
&-\frac{1}{2^{d_1-2}\pi^{\frac{d_1}{2}}}\frac{T}{a^{d_1}} \sum_{k=1}^{\infty}\sum_{l=0}^{\infty}\sum_{p=-\infty}^{\infty}e^{2\pi i k\alpha}\frac{\left(a\sqrt{\omega_{\mathcal{N};l}^2+(2\pi p T)^2+m^2}\right)^{\frac{d_1+2  }{2}}}{k^{\frac{d_1-2 }{2}}}K_{\frac{d_1+2 }{2}}\left(2ka \sqrt{\omega_{\mathcal{N};l}^2+(2\pi p T)^2+m^2}\right),
\end{split}
\end{equation*}
\begin{equation}
\label{eq2_25_1}\begin{split}\mathcal{F}_{\text{Cas}}^{\parallel}(a)=&  -\frac{d_1}{2^{d_1}\pi^{\frac{d_1+1}{2}}a^{d_1+1}}  \sum_{k=1}^{\infty} \sum_{l=0}^{\infty}e^{2\pi i k\alpha}  \left( \frac{a\sqrt{\omega_{\mathcal{N};l}^2+m^2}}{k}\right)^{\frac{d_1+1}{2}}K_{\frac{d_1+1}{2}}\left( 2ka\sqrt{\omega_{\mathcal{N};l}^2+m^2}\right)-\frac{1}{2^{d_1-1}\pi^{\frac{d_1+1}{2}}a^{d_1+1}}\\&\times \sum_{k=1}^{\infty} \sum_{l=0}^{\infty}e^{2\pi i k\alpha}
\frac{\left(a\sqrt{\omega_{\mathcal{N};l}^2+m^2}\right)^{\frac{d_1+3}{2}}}{k^{\frac{d_1-1}{2}}}K_{\frac{d_1-1}{2}}\left( 2ka\sqrt{\omega_{\mathcal{N};l}^2+m^2}\right) + \frac{(aT)^{\frac{d_1-2}{2}}}{2^{\frac{d_1-2}{2}}\pi^{\frac{d_1-4}{2}}a^{d_1+1}} \sum_{k=0}^{\infty} \sum_{l=0}^{\infty}\sum_{p=1}^{\infty} \\&\times (k+\alpha)^2  \left(\frac{ \sqrt{(\pi [k+\alpha])^2+(a\omega_{\mathcal{N};l})^2+(am)^2}}{p}\right)^{\frac{d_1-2}{2}}K_{\frac{d_1-2}{2}}\left(\frac{p}{Ta }\sqrt{(\pi [k+\alpha])^2+(a\omega_{\mathcal{N};l})^2+(am)^2}\right)\\
&-\frac{(aT)^{\frac{d_1+1}{2}}}{2^{\frac{d_1-1}{2}}\pi^{\frac{d_1+1}{2}}a^{d_1+1}}\sum_{l=0}^{\infty}\sum_{p=1}^{\infty}\left(\frac{ a\sqrt{\omega_{\mathcal{N};l}^2+m^2}}{p}\right)^{\frac{d_1+1}{2}}K_{\frac{d_1+1}{2}}\left(\frac{p}{T}\sqrt{\omega_{\mathcal{N};l}^2+m^2}\right).
\end{split}
\end{equation}The first two terms in \eqref{eq2_25_1} give the zero temperature Casimir force density and it agrees with the result obtained in \cite{2_25_1}. Notice that since the Casimir force density $\mathcal{F}_{\text{Cas}}^{\parallel}(a)$ is derived as a limit of the Casimir force $F^{\infty}_{\text{Cas}}(a)$, it follows that for a pair of infinite parallel plates with   Dirichlet or Neumann boundary conditions on both plates, the Casimir force is attractive. For a pair of infinite parallel plates with Dirichlet boundary condition on one plate and Neumann boundary condition on the other plate, the Casimir force is repulsive. For either homogeneous or mixed boundary conditions, the magnitude of the Casimir force is a decreasing function of the plate separation $a$, but it is enhanced by the presence of the extra dimensions. If both the plates assume the same (Dirichlet or Neumann) boundary conditions, the Casimir force density $\mathcal{F}_{\text{Cas}}^{\parallel}(a)$ is also a decreasing function of mass. In the high temperature regime, the leading term of the Casimir force is linear in temperature. In the low temperature regime, the Casimir effect is dominated by the zero temperature term. As is shown in Appendix \ref{a2}, the behavior of the thermal correction is rather complicated and it depends on the relative strength of $T$ and $m$. If $T\ll m$, the thermal correction term to the Casimir force density is exponentially suppressed. However, if $m\ll T\ll 1$, we find from \eqref{eq3_2_1} that the leading behavior of the thermal correction term is
\begin{equation*}
\begin{split}
-\frac{\kappa}{(4\pi)^{\frac{d_1+1}{2}}}\Biggl\{& 2^{d_1+1}\sum_{q=0}^{\frac{d_1-2}{2}}\frac{(-1)^q}{q!}\left(\frac{m}{2}\right)^{2q}\Gamma\left(\frac{1+d_1-2q}{2}\right)\zeta_R(1+d_1-2q)T^{d_1+1-2q}
\\&+\frac{(-1)^{\frac{d_1}{2}}}{\left(\frac{d_1}{2}\right)!} \sqrt{\pi} Tm^{d_1}\left(-\log\left(\frac{m}{T}\right)^2+\psi\left(\frac{d_1+2}{2}\right)-\psi(1)\right)-\frac{1}{2}\Gamma\left(-\frac{d_1+1}{2}\right)m^{d_1+1}\Biggr\}, \;\;\;\;\text{if}\;\;d_1\;\text{is even};\\
-\frac{\kappa}{(4\pi)^{\frac{d_1+1}{2}}}\Biggl\{& 2^{d_1+1}\sum_{q=0}^{\frac{d_1-1}{2}}\frac{(-1)^q}{q!}\left(\frac{m}{2}\right)^{2q}\Gamma\left(\frac{1+d_1-2q}{2}\right)\zeta_R(1+d_1-2q)T^{d_1+1-2q}
\\& +\sqrt{\pi}\Gamma\left(-\frac{d_1}{2}\right)Tm^{d_1}+\frac{1}{2}\frac{(-1)^{\frac{d_1+1}{2}}m^{d_1+1}}{\left(\frac{d_1+1}{2}\right)!}\left(\log\left(\frac{m}{4\pi T}\right)^2 -\psi\left(\frac{d_1+3}{2}\right)-\psi(1)\right)\Biggr\},\;\;\;\;\text{if}\;\;d_1\;\text{is odd}.\end{split}
\end{equation*}In the limit when the mass $m$ goes to zero, this shows that the leading order term of the temperature correction to the Casimir force density acting on a pair of parallel plates is
\begin{equation}
-\frac{\kappa}{\pi^{\frac{d_1+1}{2}}}\Gamma\left(\frac{d_1+1}{2}\right)\zeta_R(d_1+1) T^{d_1+1}.
\end{equation}This is in contrast to the case where the area of the cross section is finite, in which case the thermal correction term to the Casimir force is exponentially small when the temperature is low.

\subsection{Special Case}
Here we consider the special case as in section \ref{s3_3}, where the cross section of the cylinder $\Omega$ is a rectangular region $[0, L_2]\times \ldots\times [0, L_{d_1}]$ and the internal manifold $\mathcal{N}^n$ is an $n$-torus $T^n$ -- a product of $n$ circles with radius $r_1, \ldots, r_n$ respectively. For simplicity, we only consider the Casimir force $F_{\text{Cas}}^{\infty} (a)$ acting on a rectangular piston moving freely inside a semi-infinite long rectangular box when both the piston and the walls of the rectangular box assume the same (Dirichlet or Neumann) boundary conditions. The result for the case where they assume different boundary conditions can be obtained using \eqref{eq3_6_2}. From \eqref{eq3_4_4} and \eqref{eq2_18_2_1}, we find  that the left closed   chamber contributes the Casimir force
\begin{equation*}
\begin{split}
F_{\text{Cas}}^{\text{left}}(a; D/N) =  -\tilde{\Sigma}_1 - T \sum_{\boldsymbol{j}\in \mathbb{N}^{d_1-1}/\tilde{\mathbb{N}}^{d_1-1}}\sum_{\boldsymbol{l}\in \Z^n}\sum_{p=-\infty}^{\infty} \frac{\sqrt{\sum_{i=2}^{d_1}\left(\frac{\pi j_i}{L_i}\right)^2+ \sum_{i=1}^n \left(\frac{l_i}{r_i}\right)^2+m^2+(2\pi p T)^2}}{
\exp\left( 2a\sqrt{\sum_{i=2}^{d_1}\left(\frac{\pi j_i}{L_i}\right)^2+ \sum_{i=1}^n \left(\frac{l_i}{r_i}\right)^2+m^2+(2\pi p T)^2}\right) -1},
\end{split}
\end{equation*}and the right infinitely long chamber contributes
\begin{equation}\label{eq3_6_3}\begin{split}
F_{\text{Cas}}^{\text{right}}&( D/N) =\tilde{\Sigma}_1=-\frac{\left[\prod_{l=1}^n r_l\right]}{2^{d_1+1}}\sum_{i=0}^{d_1-1} \frac{(\mp 1)^{d_1-i-1}}{\pi^{\frac{i-n+2}{2}}}  m^{\frac{i+n+2}{2}}\sum_{2\leq \sigma_1<\ldots<\sigma_{i}\leq d_1}\left[\prod_{j=1}^i L_{\sigma_j}\right]\sum_{(k_{\sigma_1}, \ldots, k_{\sigma_i}, l_1, \ldots, l_n, p) \in \Z^{i+n+1}\setminus\{0\}}\\
& \times\left(\sum_{j=1}^i \left[L_{\sigma_j}k_{\sigma_j}\right]^2+\sum_{j=1}^n \left[\pi r_j l_j\right]^2+\left[\frac{p}{2T}\right]^2\right)^{-\frac{i+n+2}{4}}K_{\frac{i+n+2}{2}}\left( 2  m \sqrt{\sum_{j=1}^i \left[L_{\sigma_j}k_{\sigma_j}\right]^2+\sum_{j=1}^n \left[\pi r_j l_j\right]^2+\left[\frac{p}{2T}\right]^2}\right).
\end{split}\end{equation}
For Neumann boundary conditions, it is obvious that $\tilde{\Sigma}_1$ is negative. Using \eqref{eq3_4_1}, we can rewrite $\tilde{\Sigma}_1$ in the form \eqref{eq2_18_1}, where the terms with $p=0$ in \eqref{eq3_6_3} are the $T$-independent terms correspond to the first two terms in \eqref{eq2_18_1}. As is shown in Appendix \ref{a1}, when $T$ is large, $\tilde{\Sigma}_1$ is negative and dominated by a term proportional $T^{d+1}$. Therefore, when $T$ is large, there is a large repulsive force due to the vacuum fluctuations of the scalar field in the left chamber that tend to push the piston away from the closed end of the rectangular box. However, the vacuum fluctuations of the field in the right chamber give rise to a larger force in the opposite direction. The sum of  these two forces is
\begin{equation}\label{eq3_6_5}
F_{\text{Cas}}^{\infty}(a; D/N) =    - T \sum_{\boldsymbol{j}\in \mathbb{N}^{d_1-1}/\tilde{\mathbb{N}}^{d_1-1}}\sum_{\boldsymbol{l}\in \Z^n}\sum_{p=-\infty}^{\infty} \frac{\sqrt{\sum_{i=2}^{d_1}\left(\frac{\pi j_i}{L_i}\right)^2+ \sum_{i=1}^n \left(\frac{l_i}{r_i}\right)^2+m^2+(2\pi p T)^2}}{
\exp\left( 2a\sqrt{\sum_{i=2}^{d_1}\left(\frac{\pi j_i}{L_i}\right)^2+ \sum_{i=1}^n \left(\frac{l_i}{r_i}\right)^2+m^2+(2\pi p T)^2}\right) -1},
\end{equation}which tends to move the piston   towards the
closed end of the rectangular box. The expression \eqref{eq3_6_5} shows that the Casimir force $F_{\text{Cas}}^{\infty}(a; D/N)$ is exponentially small when any of the parameters $a, m, T$ is large or any of the parameters $L_2, \ldots, L_{d_1}, r_1, \ldots, r_n$ is small. There are a few alternative expressions for the Casimir force $F_{\text{Cas}}^{\infty}(a; D/N) $ which can be used to study the behaviors of the Casimir force at other limits. Using the formula \eqref{eq2_17_8}, we find that if $a$ and $ m$ are large and   $L_2, \ldots, L_{d_1}, r_1, \ldots, r_n, T$ are small, the Casimir force can be computed using the formula
\begin{equation*}
\begin{split}
&F_{\text{Cas}}^{\infty}(a; D/N)
=\frac{1}{2\pi} \sum_{k=1}^{\infty}\sum_{\boldsymbol{j}\in \mathbb{N}^{d_1-1}/\tilde{\mathbb{N}}^{d_1-1}}\sum_{\boldsymbol{l}\in \Z^n}\sum_{p=-\infty}^{\infty}  \\&\times \Biggl\{
\sqrt{\frac{ \sum_{i=2}^{d_1}\left(\frac{\pi j_i}{L_i}\right)^2+ \sum_{i=1}^n \left(\frac{l_i}{r_i}\right)^2 +m^2}{[k a]^2 +\left[\frac{p}{2T}\right]^2}} K_1\left(2\sqrt{\left(\sum_{i=2}^{d_1}\left(\frac{\pi j_i}{L_i}\right)^2+ \sum_{i=1}^n \left(\frac{l_i}{r_i}\right)^2 +m^2\right)\left([k a]^2 +\left[\frac{p}{2T}\right]^2\right)}\right)\\
&-2k^2a^2 \frac{ \sum_{i=2}^{d_1}\left(\frac{\pi j_i}{L_i}\right)^2+ \sum_{i=1}^n \left(\frac{l_i}{r_i}\right)^2 +m^2}{[k a]^2 +\left[\frac{p}{2T}\right]^2} K_2\left(2\sqrt{\left(\sum_{i=2}^{d_1}\left(\frac{\pi j_i}{L_i}\right)^2+ \sum_{i=1}^n \left(\frac{l_i}{r_i}\right)^2+m^2\right)\left([k a]^2 +\left[\frac{p}{2T}\right]^2\right)}\right)
\Biggr\}.
\end{split}
\end{equation*}In the zero temperature limit, we find that the zero temperature Casimir force is given by
\begin{equation*}
\begin{split}
&F_{\text{Cas}}^{\infty}(a; T=0; D/N)\\
=&\frac{1}{2\pi} \sum_{k=1}^{\infty}\sum_{\boldsymbol{j}\in \mathbb{N}^{d_1-1}/\tilde{\mathbb{N}}^{d_1-1}}\sum_{\boldsymbol{l}\in \Z^n}  \Biggl\{
\frac{ \sqrt{\sum_{i=2}^{d_1}\left(\frac{\pi j_i}{L_i}\right)^2+ \sum_{i=1}^n \left(\frac{l_i}{r_i}\right)^2 +m^2}}{k a } K_1\left(2 ka \sqrt{ \sum_{i=2}^{d_1}\left(\frac{\pi j_i}{L_i}\right)^2+ \sum_{i=1}^n \left(\frac{l_i}{r_i}\right)^2 +m^2  }\right)\\
&-2\left( \sum_{i=2}^{d_1}\left(\frac{\pi j_i}{L_i}\right)^2+ \sum_{i=1}^n \left(\frac{l_i}{r_i}\right)^2 +m^2\right) K_2\left(2ka\sqrt{ \sum_{i=2}^{d_1}\left(\frac{\pi j_i}{L_i}\right)^2+ \sum_{i=1}^n \left(\frac{l_i}{r_i}\right)^2+m^2 }\right)
\Biggr\}.
\end{split}
\end{equation*}This expression shows that the zero temperature Casimir force is exponentially small when any of the parameters $a, m$ is large or any of the parameters $L_2, \ldots, L_{d_1}, r_1, \ldots, r_n$ is small. For the situation that we are more interested in, i.e., the case where $r_i\ll a\ll L_j$, $1\leq i\leq n$, $2\leq j\leq d_1$, the Casimir force $F^{\infty}_{\text{Cas}}(a; D/N)$ can be written as the sum of two terms, where the first term
\begin{equation}\label{eq3_6_6}\begin{split}
&  -\frac{T}{2^{d_1-1}}\sum_{i=0}^{d_1-1} \left(\mp 1\right)^{d_1-1-i}\frac{ \tilde{S}_i }{\pi^{\frac{i+1}{2}}}\sum_{k=1}^{\infty}\sum_{\boldsymbol{l}\in\Z^n}\sum_{p=-\infty}^{\infty} \\& \times\Biggl\{i \left(\frac{
\sqrt{\sum_{j=1}^n \left(\frac{l_j}{r_j}\right)^2+ (2\pi p T)^2+m^2}}{ka}\right)^{\frac{i+1}{2}}  K_{\frac{i+1}{2}}\left(2ka\sqrt{\sum_{j=1}^n \left(\frac{l_j}{r_j}\right)^2+ (2\pi p T)^2+m^2}\right)
\\&+2\frac{\left(\sum_{j=1}^n \left(\frac{l_j}{r_j}\right)^2+ (2\pi p T)^2+m^2\right)^{\frac{i+3}{4}}}{(ka)^{\frac{i-1}{2}}} K_{\frac{i-1}{2}}\left(2ka\sqrt{\sum_{j=1}^n \left(\frac{l_j}{r_j}\right)^2+ (2\pi p T)^2+m^2}\right)
\Biggr\}\end{split}\end{equation}is the dominating term, and the second term
\begin{equation}\label{eq3_6_7}
\begin{split}
&-\frac{T}{2^{d_1}}\sum_{i=0}^{d_1-1} \frac{\left(\mp 1\right)^{d_1-1-i}}{\pi^{\frac{i+1}{2}}}\sum_{\boldsymbol{l}\in\Z^n}\sum_{p=-\infty}^{\infty} \sum_{2\leq \sigma_1<\ldots<\sigma_i\leq d_1} \left[\prod_{j=1}^{i}L_{\sigma_j}\right]
\sum_{(k_{\sigma_1}, \ldots, k_{\sigma_i})\in \Z^i\setminus\{\mathbf{0}\}} \\&\times\Biggl\{\left(\frac{ \sum_{j=1}^n \left(\frac{l_j}{r_j}\right)^2+ (2\pi p T)^2+m^2}{\sum_{j=1}^{i}[k_{\sigma_j}L_{\sigma_j} ]^2}\right)^{\frac{i+1}{4}} K_{\frac{i+1}{2}}\left(2\sqrt{\left(\sum_{j=1}^n \left(\frac{l_j}{r_j}\right)^2+ (2\pi p T)^2+m^2\right)\left(\sum_{j=1}^{i}[k_{\sigma_j}L_{\sigma_j} ]^2\right)}\right)
\\&-\frac{4\pi^{\frac{5}{2}}}{a^3}\sum_{k_1=1}^{\infty}k_1^2\left(\frac{ \left(\frac{\pi k_1}{a}\right)^2+\sum_{j=1}^n \left(\frac{l_j}{r_j}\right)^2+ (2\pi p T)^2+m^2}{\sum_{j=1}^{i}[k_{\sigma_j}L_{\sigma_j} ]^2}\right)^{\frac{i-2}{4}}\\& \times K_{\frac{i-2}{2}}\left(2\sqrt{\left(\left(\frac{\pi k_1}{a}\right)^2+\sum_{j=1}^n \left(\frac{l_j}{r_j}\right)^2+ (2\pi p T)^2+m^2\right)\left(\sum_{j=1}^{i}[k_{\sigma_j}L_{\sigma_j} ]^2\right)}\right) \Biggr\}
\end{split}\end{equation}  decays exponentially when the area of the cross section $L_2\ldots L_{d_1}$ is large.
The term $\tilde{S}_i$ in \eqref{eq3_6_6} is equal to the sum
\begin{equation*}
\tilde{S}_i=\sum_{2\leq \sigma_1<\ldots<\sigma_i\leq d_1} \left[\prod_{j=1}^{i}L_{\sigma_j}\right].
\end{equation*}The expressions \eqref{eq3_6_6} and \eqref{eq3_6_7} are suitable for investigating the high temperature behavior. The high temperature leading term is linear in $T$, which is equal to the sum of the terms  with $p=0$. At low temperature, we have the following alternative expressions for \eqref{eq3_6_6} and \eqref{eq3_6_7}:
\begin{equation}\label{eq3_6_8}\begin{split}
&  -\frac{1}{2^{d_1}}\sum_{i=0}^{d_1-1} \left(\mp 1\right)^{d_1-1-i}\frac{ \tilde{S}_i }{\pi^{\frac{i+2}{2}}}\sum_{\boldsymbol{l}\in\Z^n}  \Biggl\{-\sum_{k=1}^{\infty} \left(\frac{\sqrt{
\sum_{j=1}^n \left(\frac{l_j}{r_j}\right)^2+  m^2}}{ ka }\right)^{\frac{i+2}{2}}  K_{\frac{i+2}{2}}\left(2ka\sqrt{ \sum_{j=1}^n \left(\frac{l_j}{r_j}\right)^2 +m^2 }\right)
 \\&+2 \sum_{k=1}^{\infty}\frac{\left(
\sum_{j=1}^n \left(\frac{l_j}{r_j}\right)^2+  m^2\right)^{\frac{i+4}{4}}}{(ka)^{\frac{i}{2}} } K_{\frac{i+4}{2}}\left(2ka\sqrt{ \sum_{j=1}^n \left(\frac{l_j}{r_j}\right)^2 +m^2 }\right)+(2T)^{\frac{i+2}{2}}\sum_{p=1}^{\infty} \left(\frac{\sqrt{ \sum_{j=1}^n \left(\frac{l_j}{r_j}\right)^2+m^2}}{p}\right)^{\frac{i+2}{2}}\\&\times K_{\frac{i+2}{2}}\left(\frac{p}{T}\sqrt{\sum_{j=1}^n \left(\frac{l_j}{r_j}\right)^2+m^2}\right)-\frac{4\pi^\frac{5}{2}(2T)^{\frac{i-1}{2}}}{a^3}\sum_{p=1}^{\infty}\sum_{k=1}^{\infty}k^2\left(\frac{ \sqrt{ \left(\frac{\pi k}{a}\right)^2+\sum_{j=1}^n \left(\frac{l_j}{r_j}\right)^2+m^2}}{p}\right)^{\frac{i-1}{2}}\\& \times K_{\frac{i-1}{2}}\left(\frac{p}{T}\sqrt{ \left(\frac{\pi k}{a}\right)^2+\sum_{j=1}^n \left(\frac{l_j}{r_j}\right)^2 +m^2}\right)
\Biggr\}\end{split}\end{equation} and
\begin{equation}\label{eq3_6_9}
\begin{split}
&-\frac{1}{2^{d_1+1}}\sum_{i=0}^{d_1-1} \frac{\left(\mp 1\right)^{d_1-1-i}}{\pi^{\frac{i+2}{2}}}\sum_{\boldsymbol{l}\in\Z^n}\sum_{p=-\infty}^{\infty} \sum_{2\leq \sigma_1<\ldots<\sigma_i\leq d_1} \left[\prod_{j=1}^{i}L_{\sigma_j}\right]
\sum_{(k_{\sigma_1}, \ldots, k_{\sigma_i})\in \Z^i\setminus\{\mathbf{0}\}} \Biggl\{\left(\frac{ \sum_{j=1}^n \left(\frac{l_j}{r_j}\right)^2+m^2}{\left(\frac{p}{2T}\right)^2+\sum_{j=1}^{i}[k_{\sigma_j}L_{\sigma_j} ]^2}\right)^{\frac{i+2}{4}}\\&\times K_{\frac{i+2}{2}}\left(2\sqrt{\left(\sum_{j=1}^n \left(\frac{l_j}{r_j}\right)^2+ m^2\right)\left(\left(\frac{p}{2T}\right)^2+\sum_{j=1}^{i}[k_{\sigma_j}L_{\sigma_j} ]^2\right)}\right)
-\frac{4\pi^\frac{5}{2}}{a^3}\sum_{k_1=1}^{\infty}\left(\frac{ \left(\frac{\pi k_1}{a}\right)^2+\sum_{j=1}^n \left(\frac{l_j}{r_j}\right)^2+m^2}{\left(\frac{p}{2T}\right)^2+\sum_{j=1}^{i}[k_{\sigma_j}L_{\sigma_j} ]^2}\right)^{\frac{i-1}{4}}\\& \times K_{\frac{i-1}{2}}\left(2\sqrt{\left(\left(\frac{\pi k_1}{a}\right)^2+\sum_{j=1}^n \left(\frac{l_j}{r_j}\right)^2 +m^2\right)\left(\left(\frac{p}{2T}\right)^2+\sum_{j=1}^{i}[k_{\sigma_j}L_{\sigma_j} ]^2\right)}\right) \Biggr\}.
\end{split}\end{equation} The $p=0$ in \eqref{eq3_6_8} and \eqref{eq3_6_9} give the zero temperature Casimir force. In the limit the cross section is large, i.e. $L_i\rightarrow \infty$ for $i=2, \ldots, d_1$, we find from the term with $i=d_1-1$ in \eqref{eq3_6_6} and \eqref{eq3_6_8} that the Casimir force density acting on a pair of infinite parallel plates is given by
\begin{equation*}
\begin{split}
&\mathcal{F}_{\text{Cas}}^{\parallel}(a) \\= &  -\frac{T}{2^{d_1-1}\pi^{\frac{d_1}{2}}}    \sum_{k=1}^{\infty}\sum_{\boldsymbol{l}\in\Z^n}\sum_{p=-\infty}^{\infty} \Biggl\{(d_1-1) \left(\frac{
\sqrt{\sum_{j=1}^n \left(\frac{l_j}{r_j}\right)^2+ (2\pi p T)^2+m^2}}{ka}\right)^{\frac{d_1}{2}}  K_{\frac{d_1}{2}}\left(2ka\sqrt{\sum_{j=1}^n \left(\frac{l_j}{r_j}\right)^2+ (2\pi p T)^2+m^2}\right)
 \\&+2\frac{\left(\sum_{j=1}^n \left(\frac{l_j}{r_j}\right)^2+ (2\pi p T)^2+m^2\right)^{\frac{d_1+2}{4}}}{(ka)^{\frac{d_1-2}{2}}} K_{\frac{d_1-2}{2}}\left(2ka\sqrt{\sum_{j=1}^n \left(\frac{l_j}{r_j}\right)^2+ (2\pi p T)^2+m^2}\right)
\Biggr\},
\end{split}
\end{equation*}or alternatively
\begin{equation*}\begin{split}
& \mathcal{F}_{\text{Cas}}^{\parallel}(a) = -\frac{1}{2^{d_1}\pi^{\frac{d_1+1}{2}}}\sum_{\boldsymbol{l}\in\Z^n}  \Biggl\{- \sum_{k=1}^{\infty}\left(\frac{\sqrt{
\sum_{j=1}^n \left(\frac{l_j}{r_j}\right)^2+  m^2}}{ ka }\right)^{\frac{d_1+1}{2}}  K_{\frac{d_1+1}{2}}\left(2ka\sqrt{ \sum_{j=1}^n \left(\frac{l_j}{r_j}\right)^2 +m^2 }\right)
 \end{split}\end{equation*}\begin{equation*}\begin{split}&+2 \sum_{k=1}^{\infty}\frac{\left(
\sum_{j=1}^n \left(\frac{l_j}{r_j}\right)^2+  m^2\right)^{\frac{d_1+3}{4}}}{(ka)^{\frac{d_1-1}{2}} } K_{\frac{d_1+3}{2}}\left(2ka\sqrt{ \sum_{j=1}^n \left(\frac{l_j}{r_j}\right)^2 +m^2 }\right)+(2T)^{\frac{d_1+1}{2}}\sum_{p=1}^{\infty} \left(\frac{\sqrt{ \sum_{j=1}^n \left(\frac{l_j}{r_j}\right)^2+m^2}}{p}\right)^{\frac{d_1+1}{2}}\\&\times K_{\frac{d_1+1}{2}}\left(\frac{p}{T}\sqrt{\sum_{j=1}^n \left(\frac{l_j}{r_j}\right)^2+m^2}\right)-\frac{4\pi^\frac{5}{2}(2T)^{\frac{d_1-2}{2}}}{a^3}\sum_{p=1}^{\infty}\sum_{k=1}^{\infty}\left(\frac{ \sqrt{ \left(\frac{\pi k_1}{a}\right)^2+\sum_{j=1}^n \left(\frac{l_j}{r_j}\right)^2+m^2}}{p}\right)^{\frac{d_1-2}{2}}\\& \times K_{\frac{d_1-2}{2}}\left(\frac{p}{T}\sqrt{ \left(\frac{\pi k_1}{a}\right)^2+\sum_{j=1}^n \left(\frac{l_j}{r_j}\right)^2 +m^2}\right)
\Biggr\}.\end{split}\end{equation*}

\begin{figure}
\epsfxsize=0.5\linewidth \epsffile{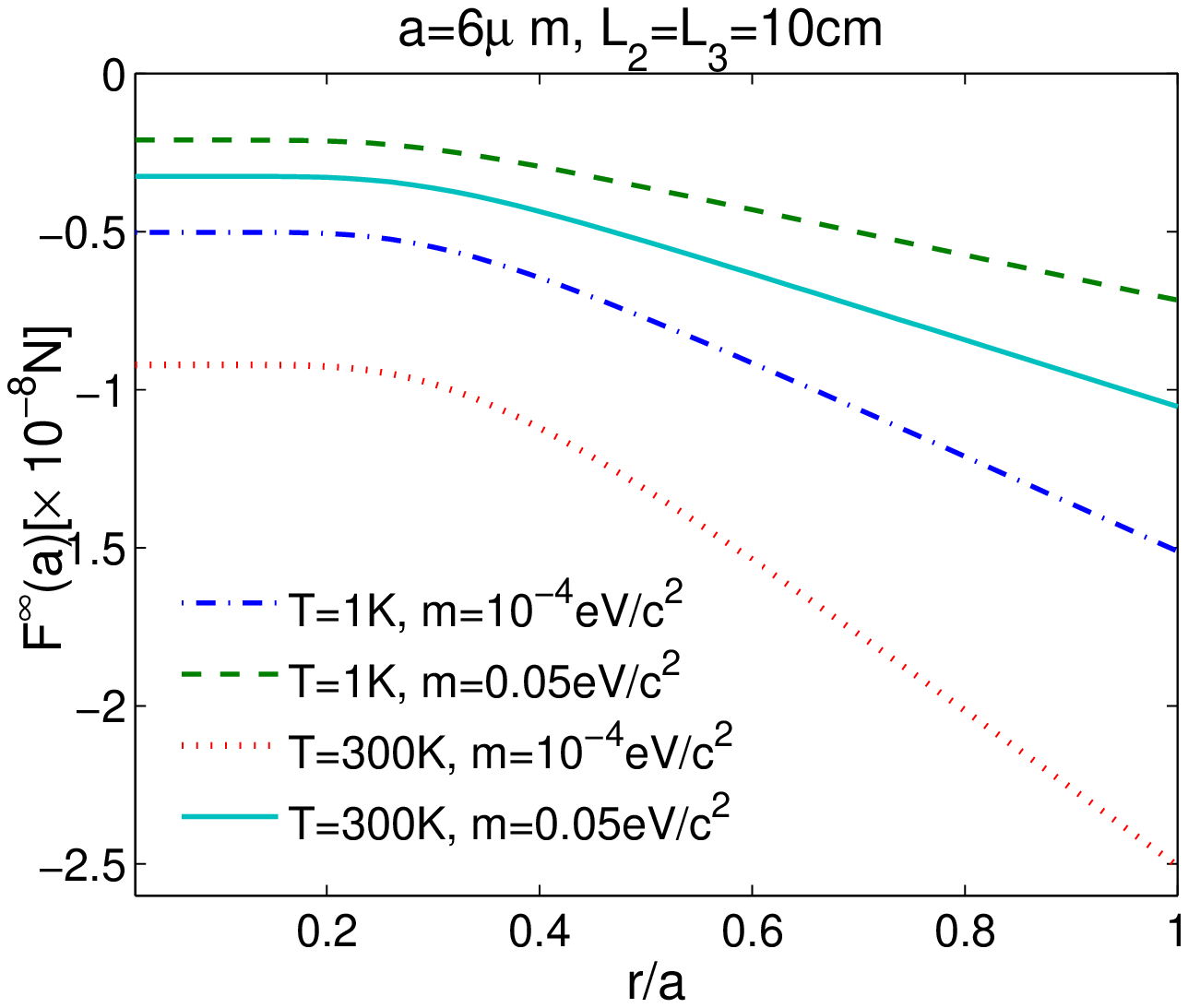} \caption{\label{f2} The Casimir force $F_{\text{Cas}}^{\infty}(a)$ as a function of $r/a$ for different values of $m$ and $T$.}\end{figure}

\begin{figure}
\epsfxsize=0.5\linewidth \epsffile{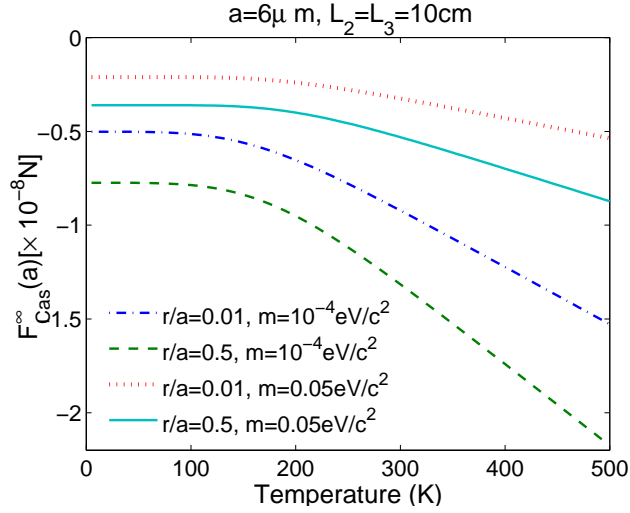} \caption{\label{f3} The Casimir force $F_{\text{Cas}}^{\infty}(a)$ as a function of temperature for different values of $r/a$ and $m$.}\end{figure}

\begin{figure}
\epsfxsize=0.5\linewidth \epsffile{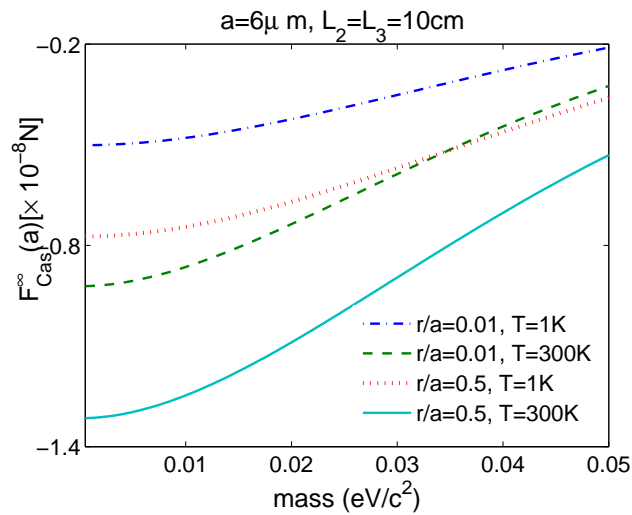} \caption{\label{f4} The Casimir force $F_{\text{Cas}}^{\infty}(a)$ as a function of mass for different values of $r/a$ and $T$.}\end{figure}

 In Figures \ref{f2}, \ref{f3} and \ref{f4}, we show graphically the behavior of the Casimir force $F_{\text{Cas}}^{\infty}(a)$ acting on two parallel plates embedded in an infinitely long rectangular cylinder in a $(4+1)$-dimensional background spacetime with an extra dimension curled up to a circle of radius $r$. We assume that the massive scalar field is subjected to Dirichlet boundary conditions on the walls of the cylinder and the two plates. Figure \ref{f2} shows the variations of the Casimir force as a function of the size $r$ of the extra dimension, or more precisely, the ratio of the size $r$ to the plate separation $a$. It shows that the magnitude of the Casimir force is increased if the size of the extra dimension is increased. When $r/a>0.4$, the existence of extra dimension can contribute substantially to the Casimir force. Figure \ref{f3} shows that the magnitude of the Casimir force is an increasing function of temperature. For $a=6\mu$m, the graph shows that the Casimir force depends linearly on temperature when $T>200K$. Figure \ref{f4} shows that the magnitude of the Casimir force decreases when the mass $m$ increases.
\section{Conclusion}
In this paper, we have investigated   the Casimir effect for massive scalar field with general curvature coupling in $(d+1)$-dimensional spacetime with $n=d-d_1$ extra dimensions. We consider the cases that the field assumes Dirichlet or Neumann boundary conditions on two parallel plates embedded in an infinitely long cylinder. We derive a general expression for the regularized Casimir energy that vanish in the infinite mass limit. A lots of the properties of the Casimir force are similar to the massless case. In particular, if the field assumes Dirichlet or Neumann boundary conditions on both plates, then the Casimir force acting on the plates are attractive. If the field assumes Dirichlet boundary condition on one plate and Neumann boundary condition on the other plate, then the Casimir force is repulsive.  Passing from massless to massive, we find that the  strength of the Casimir force is reduced if both plates assume the same boundary conditions.

  For the influence of the extra dimensions, we find that the presence of extra dimensions enhances the Casimir effect. When the size of the internal manifold shrinks to zero, one obtains the Casimir force in the $(d_1+1)$-dimensional spacetime if an only if a certain elliptic operator on the internal manifold has exactly one zero mode. This happens in particular in the case of minimal coupling or zero scalar curvature.

  For the extension of this work, it would be interesting to consider the general case of Robin boundary conditions as in \cite{2_25_1}. This will be important for considering the finite temperature Casimir effect in the braneworld model especially for the radion field stabilization mechanism, as has been shown in \cite{2_25_1} for the zero temperature case. The work along this direction will be reported elsewhere.

\begin{acknowledgments}
This project is   funded by Ministry of Science, Technology and Innovation, Malaysia under e-Science fund 06-02-01-SF0080.
\end{acknowledgments}

\appendix

\section{High temperature asymptotic behavior of $\tilde{\Sigma}_1$}\label{a1}In this section, we consider the  asymptotic behavior of $\tilde{\Sigma}_1$ \eqref{eq2_18_1} at high temperature. The first two terms in \eqref{eq2_18_1}  are independent of $T$. For the third term, we have
\begin{equation}\label{eq2_18_2}
\begin{split}
&-\frac{T}{\pi}\sum_{j\in J_*}\sum_{l=0}^{\infty}\sum_{p=1}^{\infty}
\frac{\sqrt{\omega_{\Omega,*; j}^2+\omega_{\mathcal{N};l}^2+m^2}}{p}  K_1\left(\frac{p\sqrt{\omega_{\Omega,*; j}^2+\omega_{\mathcal{N};l}^2+m^2}}{T}\right)
\\=&-\frac{1}{4\pi}\int_0^{\infty}\sum_{j\in J_*}\sum_{l=0}^{\infty}\sum_{p=1}^{\infty} \exp\left\{-\frac{\omega_{\Omega,*; j}^2+\omega_{\mathcal{N};l}^2+m^2}{t}-\frac{tp^2}{4T^2}\right\}dt\\
=&-\frac{1}{4\pi}\int_0^{\infty} \sum_{p=1}^{\infty} \exp\left\{-\frac{tp^2}{4T^2}\right\} \frac{1}{2\pi i}\int_{\text{c}-i\infty}^{\text{c}+i\infty} \Gamma(z) t^z \zeta_{\Omega\times\mathcal{N}, *}(z;m) dz dt\\
=&-\frac{1}{4\pi}
 \frac{1}{2\pi i}\int_{\text{c}-i\infty}^{\text{c}+i\infty} \Gamma(z)  \Gamma(z+1)(2T)^{2z+2} \zeta_R(2z+2) \zeta_{\Omega\times\mathcal{N}, *}(z;m) dz \\
 \sim &-\frac{1}{4\pi}\sum_{i=0}^{d-1} \Gamma\left(\frac{i+2}{2}\right)\zeta_R(i+2) c_{\Omega\times\mathcal{N},*;d-1-i}(m)(2T)^{i+2}+O(T).
\end{split}
\end{equation}

\section{The asymptotic behavior of the Casimir force $F^{\infty}_{\text{Cas}}(a)$ in different limits} \label{a2}In this section, we derive the asymptotic behavior of the Casimir force $F^{\infty}_{\text{Cas}}(a)$ in different limits. We only consider the case where $\alpha=0$ or $1$. The case $\alpha=1/2$ can be derived analogously.

First we consider the asymptotic behavior when $r<a<R$. In the high temperature regime, it follows from \eqref{eq2_16_13} that
\begin{equation}\label{eq2_19_4}
\begin{split}
 F^{\infty}_{\text{Cas}}(a)=&
 \frac{T}{2\sqrt{\pi}}\frac{\pa}{\pa a}\left\{ a\int_0^{\infty} t^{-\frac{1}{2}} \frac{1}{2\pi i}\int_{\text{c}-i\infty}^{\text{c}+i\infty}
\Gamma(z) t^z \zeta_{\Omega, *}(z) \sum_{k=1}^{\infty}\sum_{l=0}^{\infty}\sum_{p=-\infty}^{\infty} \exp\left( -\frac{ \omega_{\mathcal{N};l}^2+(2\pi p T)^2 +m^2}{t}-tk^2a^2\right) dz dt\right\}\\\sim &\frac{T}{2\sqrt{\pi}}\sum_{i=0}^{\infty}c_{\Omega,*;i}\frac{\pa}{\pa a}\left\{a\int_0^{\infty} t^{\frac{d_1-i}{2}-1}\sum_{k=1}^{\infty}\sum_{l=0}^{\infty}\sum_{p=-\infty}^{\infty}  \exp\left( -\frac{ \omega_{\mathcal{N};l}^2+(2\pi p T)^2 +m^2}{t}-tk^2a^2\right)  dt\right\},
 \end{split}\end{equation} where $\zeta_{\Omega, *}(z)$ is the zeta function
 \begin{equation*}
 \zeta_{\Omega, *}(s)=\sum_{j\in J_*} \omega_{\Omega, *, j}^{-2s},
 \end{equation*}
 and
 $c_{\Omega,*;i}$ is the heat kernel coefficient of the Laplace operator with Dirichlet/Neumann boundary conditions on $\Omega$, i.e.,
\begin{equation*}
\sum_{j\in J_*} e^{-t\omega_{\Omega,*;j}^2} = \sum_{i=0}^{M-1}c_{\Omega,*;i}t^{\frac{i-d_1+1}{2}}+O\left(t^{\frac{M-d_1+1}{2}}\right)\hspace{1cm}\text{as}\;\;t\rightarrow 0^+.
\end{equation*}
In terms of the measure $R$ of the size of $\Omega$, $$c_{\Omega, *; i}=R^{d_1-i-1}c_{\Omega/R, *;i} \propto R^{d_1-i-1}.$$ Therefore when $r<a<R$,
\begin{equation*}
\begin{split}
&F^{\infty}_{\text{Cas}}(a)\\\sim &\frac{T}{a\sqrt{\pi}}\sum_{i=0}^{M-1}c_{\Omega/R,*;i} \left(\frac{R}{a}\right)^{d_1-i-1}\sum_{k=1}^{\infty}\sum_{l=0}^{\infty}\sum_{p=-\infty}^{\infty}\Biggl\{ \left(\frac{a\sqrt{\omega_{\mathcal{N};l}^2+(2\pi p T)^2+m^2}}{k}\right)^{\frac{d_1 -i}{2}}K_{\frac{d_1-i }{2}}\left(2ka \sqrt{\omega_{\mathcal{N};l}^2+(2\pi p T)^2+m^2}\right)
\end{split}\end{equation*}\begin{equation}\label{eq2_23_1}\begin{split}&-2\frac{\left(a\sqrt{\omega_{\mathcal{N};l}^2+(2\pi p T)^2+m^2}\right)^{\frac{d_1+2-i }{2}}}{k^{\frac{d_1-2-i }{2}}}K_{\frac{d_1+2-i }{2}}\left(2ka \sqrt{\omega_{\mathcal{N};l}^2+(2\pi p T)^2+m^2}\right)\Biggr\}+O\left( \left(\frac{R}{a}\right)^{d_1-M-1}\right).
\end{split}
\end{equation}
This expansion only gives the behavior when $am\gg 1$. For the expansion when  $am\ll 1$, we go back to \eqref{eq2_19_4} and consider
 the expansion of
\begin{equation}\label{eq2_20_4}\mathfrak{T}(i;a)=\frac{\pa}{\pa a}\left\{a\int_0^{\infty} t^{\frac{d_1-i}{2}-1}\sum_{k=1}^{\infty}\sum_{l=0}^{\infty}\sum_{p=-\infty}^{\infty}  \exp\left( -\frac{ \omega_{\mathcal{N};l}^2+(2\pi p T)^2 +m^2}{t}-tk^2a^2\right)  dt\right\}\end{equation}when $am\ll 1 $.  Recall that there are $\kappa$ zero eigenvalues $\omega_{\mathcal{N};l}^2$. We  write \eqref{eq2_20_4} as the sum of $\mathfrak{T}_1(i;a)$ and $\mathfrak{T}_2(i;a)$, where
\begin{equation*}
\mathfrak{T}_1(i;a)=\kappa\frac{\pa}{\pa a}\left\{a\int_0^{\infty} t^{-\frac{i+2-d_1}{2}}\sum_{k=1}^{\infty}   \exp\left( -\frac{ m^2}{t}-tk^2a^2\right)  dt\right\}
\end{equation*}and
\begin{equation*} \mathfrak{T}_2(i;a)=\frac{\pa}{\pa a}\left\{a\int_0^{\infty} t^{\frac{d_1-i}{2}-1}\sum_{k=1}^{\infty}\sum_{\substack{(l,p)\in \tilde{\mathbb{N}}\times\Z\\
\omega_{\mathcal{N};l}^2+(2\pi pT)^2\neq 0}}  \exp\left( -\frac{ \omega_{\mathcal{N};l}^2+(2\pi p T)^2 +m^2}{t}-tk^2a^2\right)  dt\right\}.\end{equation*}
For the term $\mathfrak{T}_2(i;a)$, we can use the Taylor expansion of $e^{-m^2/t}$ to write
\begin{equation*}
\begin{split}
\mathfrak{T}_2(i;a)=&\sum_{q=0}^{\infty}\frac{(-1)^q}{q!}m^{2q}\frac{\pa}{\pa a}\left\{a\int_0^{\infty} t^{\frac{d_1-i}{2}-q-1}\sum_{k=1}^{\infty}\sum_{\substack{(l,p)\in \tilde{\mathbb{N}}\times\Z\\
\omega_{\mathcal{N};l}^2+(2\pi pT)^2\neq 0}}  \exp\left( -\frac{ \omega_{\mathcal{N};l}^2+(2\pi p T)^2  }{t}-tk^2a^2\right)  dt\right\}\\
=&2a^{i-d_1}\sum_{q=0}^{\infty}\frac{(-1)^q}{q!}(am)^{2q}\Biggl\{ \left(\frac{a\sqrt{\omega_{\mathcal{N};l}^2+(2\pi p T)^2}}{k}\right)^{\frac{d_1-i-2q}{2}}K_{\frac{d_1-i-2q}{2}}\left(2ka \sqrt{\omega_{\mathcal{N};l}^2+(2\pi p T)^2}\right)\\
&-2\frac{\left(a\sqrt{\omega_{\mathcal{N};l}^2+(2\pi p T)^2}\right)^{\frac{d_1+2-i-2q}{2}}}{(k)^{\frac{d_1-2-i-2q}{2}}}K_{\frac{d_1+2-i-2q}{2}}\left(2ka \sqrt{\omega_{\mathcal{N};l}^2+(2\pi p T)^2}\right)\Biggr\}.
\end{split}
\end{equation*} For $\mathfrak{T}_1(i;a)$, notice that when $s$ is large enough, and $am\ll 1$,
\begin{equation*}\begin{split}
\mathfrak{P}(s;a)=&\int_0^{\infty} t^{-s}\sum_{k=1}^{\infty}   \exp\left( -\frac{ m^2}{t}-tk^2a^2\right)  dt=\int_0^{\infty} t^{s-2}\sum_{k=1}^{\infty}   \exp\left( -t m^2 -\frac{k^2a^2}{t}\right)  dt
\\=& -\frac{1}{2}\Gamma(s-1)m^{-2s+2}+\frac{\sqrt{\pi}}{2a} \int_0^{\infty} t^{s-\frac{3}{2}}\sum_{k=-\infty}^{\infty}   \exp\left( -t m^2 -\frac{t \pi^2 k^2 }{a^2}\right)  dt\\
=&-\frac{1}{2}\Gamma(s-1)m^{-2s+2}+\frac{\sqrt{\pi}}{2a}\Gamma\left(s-\frac{1}{2}\right)m^{-2s+1} +\frac{\sqrt{\pi}}{a}\sum_{q=0}^{\infty}\frac{(-1)^q}{q!}m^{2q} \Gamma\left(s-\frac{1}{2}+q\right) \zeta_R(2s-1+2q)\left(\frac{\pi}{a}\right)^{-2s+1-2q}.
\end{split}\end{equation*}By analytic continuation, this formula holds for all $s$. If $s>1$, we can put in directly the value of $s$ in each of the terms. For $s\leq 1$, we have for $j= 0, 1, 2, \ldots$,
\begin{equation*}\begin{split}
\mathfrak{P}\left(1-j;a\right)=&\frac{1}{2}\frac{(-1)^{j}}{j!}m^{2j}\left(\log\left(\frac{am}{2\pi}\right)^2-\psi\left(j+1\right)-\psi(1)\right) +\frac{\sqrt{\pi}}{2a}\Gamma\left(-j+\frac{1}{2}\right)m^{2j-1} \\&+\frac{\sqrt{\pi}}{a}\sum_{\substack{q\in\tilde{\mathbb{N}}\\q\neq j}}\frac{(-1)^q}{q!}m^{2q} \Gamma\left(-j+\frac{1}{2}+q\right) \zeta_R(-2j+1+2q)\left(\frac{\pi}{a}\right)^{2j-1-2q},
\end{split}
\end{equation*}
\begin{equation*}\begin{split}
\mathfrak{P}\left(\frac{1}{2}-j;a\right)=&\frac{\sqrt{\pi}}{2a}\frac{(-1)^{j}}{j!}m^{2j}\left(-\log\left(2am \right)^2+\psi\left(j+1\right)-\psi(1)\right) -\frac{1}{2}\Gamma\left(-j-\frac{1}{2}\right)m^{2j+1} \\&+\frac{\sqrt{\pi}}{a}\sum_{\substack{q\in\tilde{\mathbb{N}}\\q\neq j}}\frac{(-1)^q}{q!}m^{2q} \Gamma\left(-j+q\right) \zeta_R(-2j+2q)\left(\frac{\pi}{a}\right)^{2j-2q}.
\end{split}
\end{equation*}Therefore, when $r<a<R$ and $am \ll 1\ll Rm$, the Casimir force $F^{\infty}_{\text{Cas}}(a)$ has the asymptotic expansion
\begin{equation}\label{eq2_23_5}
\begin{split}
& F^{\infty}_{\text{Cas}}(a)\sim  \frac{\kappa T}{2\sqrt{\pi}a}\sum_{j=0}^{\left[\frac{d_1}{2}\right]}c_{\Omega/R,*;d_1-2j}\left(\frac{R } {a}\right)^{2j-1} \Biggl\{\frac{1}{2}\frac{(-1)^{j}}{j!}(am)^{2j}\left(\log\left(\frac{am}{2\pi}\right)^2+2-\psi\left(j+1\right)-\psi(1)\right) + \sqrt{\pi} \sum_{\substack{q\in\tilde{\mathbb{N}}\\q\neq j}}\frac{(-1)^q}{q!}(am)^{2q}\\&\times (2q-2j+1)\pi^{2j-2q-1} \Gamma\left(-j+\frac{1}{2}+q\right) \zeta_R(-2j+1+2q) \Biggr\}  + \frac{\kappa T}{2\sqrt{\pi}a}\sum_{j=0}^{\left[\frac{d_1-1}{2}\right]} c_{\Omega/R,*;d_1-2j-1} \left(\frac{R}{a}\right)^{2j} \Biggl\{ -\sqrt{\pi}\frac{(-1)^{j}}{j!}(am)^{2j}\\&-\frac{1}{2}\Gamma\left(-j-\frac{1}{2}\right)(am)^{2j+1}   + \sqrt{\pi} \sum_{\substack{q\in\tilde{\mathbb{N}}\\q\neq j}}\frac{(-1)^q}{q!}(am)^{2q}(2q-2j)\pi^{2j-2q} \Gamma\left(-j+q\right) \zeta_R(-2j+2q) \Biggr\}
 + \frac{\kappa T}{2\sqrt{\pi} a} \sum_{i=d_1+1}^{\infty}\\&\times c_{\Omega/R,*;i} \left( \frac{R}{a}\right)^{d_1-1-i}\left\{ -\frac{1}{2}\frac{\Gamma\left(\frac{i-d_1}{2}\right)}{(am)^{i-d_1}}+\sqrt{\pi}
\sum_{q=0}^{\infty}\frac{(-1)^q}{q!}(am)^{2q}\frac{(i-d_1+1+2q) }{\pi^{i-d_1+1+2q}}\Gamma\left(\frac{ i-d_1+1+2q}{2}\right)\zeta_R\left(i-d_1+1+2q\right)\right\}\\
&+\frac{T}{\sqrt{\pi} a}\sum_{i=0}^{\infty} c_{\Omega,*;i}\left(\frac{R}{ a}\right)^{d_1-1-i}\sum_{q=0}^{\infty}\frac{(-1)^q}{q!}(am)^{2q}\sum_{k=1}^{\infty}\sum_{\substack{(l,p)\in \tilde{\mathbb{N}}\times\Z\\
\omega_{\mathcal{N};l}^2+(2\pi pT)^2\neq 0}}\Biggl\{ \left(\frac{a\sqrt{\omega_{\mathcal{N};l}^2+(2\pi p T)^2}}{k}\right)^{\frac{d_1-i-2q}{2}}\\&\times K_{\frac{d_1-i-2q}{2}}\left(2ka \sqrt{\omega_{\mathcal{N};l}^2+(2\pi p T)^2}\right)-2\frac{\left(a\sqrt{\omega_{\mathcal{N};l}^2+(2\pi p T)^2}\right)^{\frac{d_1+2-i-2q}{2}}}{k^{\frac{d_1-2-i-2q}{2}}}K_{\frac{d_1+2-i-2q}{2}}\left(2ka \sqrt{\omega_{\mathcal{N};l}^2+(2\pi p T)^2}\right)\Biggr\}.
\end{split}
\end{equation}

In the low temperature regime, we use \eqref{eq2_10_5} to transform  \eqref{eq2_17_8} to
\begin{equation*}\begin{split}
F^{\infty}_{\text{Cas}}(a)=&\frac{1}{4\pi}\frac{\pa}{\pa a}\Biggl\{ a\int_0^{\infty}  \sum_{k=1}^{\infty}\sum_{j\in J_*}\sum_{l=0}^{\infty} \exp\left\{-\frac{ \omega_{\Omega,*;j}^2+\omega_{\mathcal{N}, l}^2 +m^2}{t}-t k^2 a^2  \right\}dt\\
&+2\sqrt{\pi}\int_0^{\infty} t^{-\frac{1}{2}} \sum_{k=1}^{\infty}\sum_{j\in J_*}\sum_{l=0}^{\infty}\sum_{p=1}^{\infty} \exp\left\{-\frac{1}{t}\left( \left[\frac{\pi k}{a}\right]^2+ \omega_{\Omega,*;j}^2+\omega_{\mathcal{N}, l}^2 +m^2 \right)-t \left(\frac{p}{2T}\right)^2 \right\}dt\Biggr\}\\
&-\frac{1}{4\pi}\int_0^{\infty}  \sum_{j\in J_*}\sum_{l=0}^{\infty}\sum_{p=1}^{\infty} \exp\left\{-\frac{ \omega_{\Omega,*;j}^2+\omega_{\mathcal{N}, l}^2 +m^2}{t}-t \left(\frac{p}{2T}\right)^2 \right\}dt.
\end{split}\end{equation*}Using the same method as we derive \eqref{eq2_23_1}, we find that
\begin{equation*}
\begin{split}
&F^{\infty}_{\text{Cas}}(a)\sim \sum_{i=0}^{\infty} c_{\Omega/R, *; i}R^{d_1-1-i} \Bigg\{ \frac{1}{4\pi}\frac{\pa}{\pa a}\Biggl[ a\int_0^{\infty}  t^{\frac{d_1-1-i}{2}} \sum_{k=1}^{\infty} \sum_{l=0}^{\infty} \exp\left\{-\frac{  \omega_{\mathcal{N}, l}^2 +m^2}{t}-t k^2 a^2  \right\}dt\\
&+2 \sqrt{\pi} \int_0^{\infty} t^{\frac{d_1-i-2}{2}} \sum_{k=1}^{\infty} \sum_{l=0}^{\infty}\sum_{p=1}^{\infty} \exp\left\{-\frac{1}{t}\left( \left[\frac{\pi k}{a}\right]^2 +\omega_{\mathcal{N}, l}^2 +m^2 \right)-t \left(\frac{p}{2T}\right)^2 \right\}dt\Biggr]\\
&-\frac{1}{4\pi}\int_0^{\infty}  t^{\frac{d_1-1-i}{2}}  \sum_{l=0}^{\infty}\sum_{p=1}^{\infty} \exp\left\{-\frac{  \omega_{\mathcal{N}, l}^2 +m^2}{t}-t \left(\frac{p}{2T}\right)^2 \right\}dt\Biggr\}\\
\end{split}\end{equation*}\begin{equation}\label{eq2_23_3}\begin{split}=& -\frac{1}{a^2}\sum_{i=0}^{\infty} c_{\Omega/R, *; i}\left(\frac{R}{a}\right)^{d_1-1-i} \Biggl\{ \frac{1}{2\pi}\sum_{k=1}^{\infty} \sum_{l=0}^{\infty}\Biggl[ (d_1-i) \left( \frac{a\sqrt{\omega_{\mathcal{N};l}^2+m^2}}{k}\right)^{\frac{d_1+1-i}{2}}K_{\frac{d_1+1-i}{2}}\left( 2ka\sqrt{\omega_{\mathcal{N};l}^2+m^2}\right)\\&+2
\frac{\left(a\sqrt{\omega_{\mathcal{N};l}^2+m^2}\right)^{\frac{d_1+3-i}{2}}}{k^{\frac{d_1-1-i}{2}}}K_{\frac{d_1-1-i}{2}}\left( 2ka\sqrt{\omega_{\mathcal{N};l}^2+m^2}\right)\Biggr]- 2^{\frac{d_1-i}{2}}\pi^{\frac{3}{2}}(aT)^{\frac{d_1-i-2}{2}} \sum_{k=1}^{\infty} \sum_{l=0}^{\infty}\sum_{p=1}^{\infty} k^2 \\&\times \left(\frac{ \sqrt{(\pi k)^2+(a\omega_{\mathcal{N};l})^2+(am)^2}}{p}\right)^{\frac{d_1-i-2}{2}}K_{\frac{d_1-i-2}{2}}\left(\frac{p}{Ta }\sqrt{(\pi k)^2+(a\omega_{\mathcal{N};l})^2+(am)^2}\right)\\
&+\frac{(2aT)^{\frac{d_1-i+1}{2}}}{2\pi}\sum_{l=0}^{\infty}\sum_{p=1}^{\infty}\left(\frac{ a\sqrt{\omega_{\mathcal{N};l}^2+m^2}}{p}\right)^{\frac{d_1-i+1}{2}}K_{\frac{d_1-i+1}{2}}\left(\frac{p}{T}\sqrt{\omega_{\mathcal{N};l}^2+m^2}\right)\Biggr\}.
\end{split}
\end{equation}In case $T=0$ and $am\ll 1 \ll Rm$,
\begin{equation}\label{eq2_26_1}
\begin{split}
& F^{\infty, T=0}_{\text{Cas}}(a)\\\sim & \frac{ \kappa}{4\pi a^2}\sum_{j=0}^{\left[\frac{d_1+1}{2}\right]}c_{\Omega/R,*;d_1+1-2j}\left(\frac{R } {a}\right)^{2j-2} \Biggl\{\frac{1}{2}\frac{(-1)^{j}}{j!}(am)^{2j}\left(\log\left(\frac{am}{2\pi}\right)^2+2-\psi\left(j+1\right)-\psi(1)\right) + \sqrt{\pi} \sum_{\substack{q\in\tilde{\mathbb{N}}\\q\neq j}}\frac{(-1)^q}{q!}(am)^{2q}\\&\times (2q-2j+1)\pi^{2j-2q-1} \Gamma\left(-j+\frac{1}{2}+q\right) \zeta_R(-2j+1+2q) \Biggr\}  +\frac{\kappa }{4\pi a^2}\sum_{j=0}^{\left[\frac{d_1}{2}\right]} c_{\Omega/R,*;d_1-2j} \left(\frac{R}{a}\right)^{2j-1} \Biggl\{ -\sqrt{\pi}\frac{(-1)^{j}}{j!}(am)^{2j}\\&-\frac{1}{2}\Gamma\left(-j-\frac{1}{2}\right)(am)^{2j+1}   + \sqrt{\pi} \sum_{\substack{q\in\tilde{\mathbb{N}}\\q\neq j}}\frac{(-1)^q}{q!}(am)^{2q}(2q-2j)\pi^{2j-2q} \Gamma\left(-j+q\right) \zeta_R(-2j+2q) \Biggr\}
 + \frac{\kappa }{4\pi a^2} \sum_{i=d_1+2}^{\infty}\\&\times c_{\Omega/R,*;i} \left( \frac{R}{a}\right)^{d_1-1-i}\left\{ -\frac{1}{2}\frac{\Gamma\left(\frac{i-1-d_1}{2}\right)}{(am)^{i-1-d_1}}+\sqrt{\pi}
\sum_{q=0}^{\infty}\frac{(-1)^q}{q!}(am)^{2q}\frac{(i-d_1+2q) }{\pi^{i-d_1+2q}}\Gamma\left(\frac{ i-d_1+2q}{2}\right)\zeta_R\left(i-d_1+2q\right)\right\}\\
&-\frac{1}{2\pi a^2}\sum_{ i=0}^{\infty} c_{\Omega,*;i}\left(\frac{R}{ a}\right)^{d_1-1-i}\sum_{q=0}^{\infty}\frac{(-1)^q}{q!}(am)^{2q}\sum_{k=1}^{\infty}\sum_{\substack{l\in \tilde{\mathbb{N}}\\ \omega_{\mathcal{N},l}^2\neq 0}}\Biggl\{ (d_1-i-2q) \left(\frac{\omega_{\mathcal{N};l}}{k}\right)^{\frac{d_1+1-i-2q}{2}}K_{\frac{d_1+1-i-2q}{2}}\left(2ka\omega_{\mathcal{N};l}\right)\\
&+2\frac{\omega_{\mathcal{N};l}^{\frac{d_1+3-i-2q}{2}}}{k^{\frac{d_1-1-i-2q}{2}}}K_{\frac{d_1-1-i-2q}{2}}\left(2ka\omega_{\mathcal{N};l}\right)\Biggr\}.
\end{split}
\end{equation}For the temperature correction terms, it can also be written as a power series in $m^2$ if $T<m$. However, if $m \ll T$, we have a contribution
\begin{equation*}
\begin{split}
&-\frac{\kappa m^2 }{4\pi }\sum_{j=0}^{\left[\frac{d_1+1}{2}\right]}c_{\Omega/R,*;d_1+1-2j} (Rm)^{2j-2} \Biggl\{ \frac{1}{2}\frac{(-1)^{j}}{j!} \left(\log\left(\frac{m}{4\pi T}\right)^2-\psi\left(j+1\right)-\psi(1)\right) +\sqrt{\pi} \Gamma\left(-j+\frac{1}{2}\right)\frac{T}{m} \\&+ 2\sqrt{\pi} \sum_{\substack{q\in\tilde{\mathbb{N}}\\q\neq j}}\frac{(-1)^q}{q!}  \Gamma\left(-j+\frac{1}{2}+q\right) \zeta_R(-2j+1+2q)\left(2\pi \right)^{2j-1-2q}\left(\frac{m}{T}\right)^{2q-2j}\Biggr\}-\frac{\kappa m^2 }{4\pi }\sum_{j=0}^{\left[\frac{d_1}{2}\right]}c_{\Omega/R,*;d_1-2j} (Rm)^{2j-1} \\&\times \Biggl\{ \sqrt{\pi}\frac{(-1)^{j}}{j!}\frac{T}{m}\left(-\log\left(\frac{m}{T} \right)^2+\psi\left(j+1\right)-\psi(1)\right) -\frac{1}{2}\Gamma\left(-j-\frac{1}{2}\right)   +2\sqrt{\pi} \sum_{\substack{q\in\tilde{\mathbb{N}}\\q\neq j}}\frac{(-1)^q}{q!}\end{split}\end{equation*}\begin{equation}\label{eq3_2_1}\begin{split}&\times \Gamma\left(-j+q\right) \zeta_R(-2j+2q)\left(2\pi \right)^{2j-2q}\left(\frac{m}{T}\right)^{2q-2j-1}\Biggr\}
-\frac{\kappa m^2}{4\pi}\sum_{i=d_1+2}^{\infty}c_{\Omega/R, *; i} (Rm)^{d_1-1-i} \Biggl\{ -\frac{1}{2}\Gamma\left(\frac{i-d_1-1}{2}\right) \\&+ \sqrt{\pi}\Gamma\left(\frac{i-d_1}{2}\right)\frac{T}{m} +2\sqrt{\pi}\sum_{q=0}^{\infty}\frac{(-1)^q}{q!}  \Gamma\left(\frac{i-d_1+2q}{2}\right) \zeta_R(i-d_1+2q)\left(2\pi \right)^{d_1-i-2q}\left(\frac{m}{T}\right)^{i-d_1+2q-1}\Biggr\}
\end{split}
\end{equation} coming from the terms with $\omega_{\mathcal{N};l}^2=0$ in the last term of \eqref{eq2_23_3}. It contains logarithmic terms of $m/T$.

 \end{document}